\documentclass[aps,superscriptaddress,nofootinbib,notitlepage,11pt,tightenlines]{revtex4-1}

\usepackage{mathptmx}       % selects Times Roman as basic font
\usepackage{helvet}         % selects Helvetica as sans-serif font
\usepackage{courier}        % selects Courier as typewriter font

\usepackage[bottom]{footmisc}% places footnotes at page bottom

\usepackage[pdftex]{graphicx}

\usepackage{amsmath}
\usepackage{bm}
\usepackage{hyperref}

\hypersetup
{%
pdftitle = {Photon-Atom Coupling with Parabolic Mirrors},
pdfauthor = {Markus Sondermann and Gerd Leuchs},
colorlinks = {true},
urlcolor=blue,
linkcolor=blue
}

\begin{document}
\title{Photon-Atom Coupling with Parabolic Mirrors}

\author{Markus Sondermann}
\email{markus.sondermann@fau.de} 
\affiliation{University of Erlangen-Nuremberg,
  Department of Physics, Staudtstr. 7/B2, 91058 Erlangen, Germany}

\author{Gerd Leuchs}
\email{gerd.leuchs@mpl.mpg.de}
\affiliation{University of Erlangen-Nuremberg,
  Department of Physics, Staudtstr. 7/B2, 91058 Erlangen, Germany}
\affiliation{Max Planck Institute for the Science of Light,
Guenther-Scharowsky-Str. 1/Bldg. 24, 91058 Erlangen, Germany}
\affiliation{Department of Physics, University of Ottawa, Ottawa,
Ont. K1N 6N5, Canada}

\begin{abstract}
Efficient coupling of light to single atomic systems has gained
considerable attention over the past decades. 
This development is driven by the continuous growth of quantum
technologies.
The efficient coupling of light
and matter is an enabling technology for quantum information processing and
quantum communication. 
And indeed, in recent years much progress has been made in this
direction.
But applications aside, the interaction of photons and atoms is a
fundamental physics problem.
There are various possibilities for making this interaction more
efficient, among them the apparently \lq natural\rq\ attempt of mode-matching
the light field to the free-space emission pattern of the atomic
system of interest. 
Here we will describe the necessary steps of implementing this
mode-matching with the ultimate aim of reaching
unit coupling efficiency.
We describe the use of deep parabolic mirrors as the central optical
element of a free-space coupling scheme, covering the preparation of
suitable modes of the field incident onto these mirrors as well as the
location of an atom at the mirror's focus.
Furthermore, we establish a robust method for determining the
efficiency of the photon-atom coupling.
\end{abstract}

\maketitle

%%%%%%%%%%%%%%%%%%%%%%%%%%%%%%%%%%%%%%%%%%%%%%%
\section{Coupling to an atom: the role of dipole radiation}
\label{sec.dipole}

\subsection{General considerations}

When discussing the interaction of photons and atoms it is
a good idea to make some simplifying approximations. 
First, we neglect the complexity of nature and assume that the atom
has only two energy levels.
Then, we take advantage of the fact that the extent of the atom in
real space is much smaller than the wavelength of the light emitted by
the atom and make the so-called \emph{dipole
  approximation}~\cite{mandel-wolf1995,scully1997}.
The dipole approximation allows us to treat the atom as an electric
dipole while neglecting magnetic dipoles and higher-order multipoles.  
Within this approximation the interaction energy of the atom and the light
field is given by the scalar product of the electric field vector at the
position of the atom and the electric dipole moment of the atomic
transition:
\begin{equation}
H_I= -\bm{\mu}\cdot\mathbf{E}\quad .
\end{equation}
This suggests to put as much as possible of the incident electric
field into the vector component parallel to the atomic
dipole in order to maximize the interaction energy.
Despite of some applications such as laser cooling where one can
just put more power into the incident beam to adjust the interaction
strength, more sophisticated methods are mandatory when for example 
coupling a single photon to an atom, where the amount of available
energy is naturally limited.\footnote{There is no state of the light
  field, more intense than a 
  single photon, which will transform into a single photon state when
projecting onto a certain mode function. Such a projection is
equivalent to attenuation.}

There are various strategies for enhancing the interaction of photons
and atoms.\footnote{If not explicitly mentioned otherwise, the term
  \emph{atom} is used to designate any kind of single quantum target.}
One is to enhance the field strength by placing the atom in the near
field of a suitable antenna, which can enhance the local field strength
far above the value given by the diffraction limited focusing of the
incident field in the absence of the
antenna~\cite{moskovits1985,kuehn2006,novotny2011}.
The approach followed most frequently in the past decades is to place
the atom in an anti-node of the electric field of a high quality
resonator (see Refs.~\cite{raimond2001, walther2006, kimble1998,
  rempe1993} for reviews and also the chapters by
\href{http://arxiv.org/abs/1502.01062}{Lanco \&   Senellart} 
and \href{http://arxiv.org/abs/1502.06741}{A.~Kuhn}).
It is the small cavity mode volume and the fact that the light
field interacts with the atom over many cavity
round-trips~\cite{sondermann2007} that enhances the interaction.
Recently, impressive progress in the field of so-called
\emph{cavity quantum-electrodynamics} was achieved (see
e.g. Refs.~\cite{specht2011,ritter2012,sames2014,tiecke2014}).

This chapter (and also the ones by
\href{http://arxiv.org/abs/1502.04302}{Slodicka et al} and 
\href{http://arxiv.org/abs/1502.04349}{Piro \& Eschner}) is devoted to
the coupling of atoms and the light field in \emph{free space}.
Contrary to the approaches mentioned above, efficient free-space
photon-atom coupling is not based on modifying the boundary conditions
of the electro-magnetic field but rather on suitably shaping the field
itself.
Remember the dipole approximation mentioned above. It is the very name
of this approximation that suggests how to couple light efficiently to
the atom: shape the incident field as to resemble the kind of dipole
radiation emitted by the atom! 
This mode-matching argument is basically the same as when thinking
of coupling light efficiently into a single-mode optical fibre
and follows directly from time-reversal symmetry
arguments (see
Refs.~\cite{quabis2000,leuchs2012,leuchs2013o,silberfarb2003,pinotsi2008}
 and references therein).

\subsection{Defining a coupling efficiency}

Thinking about coupling in a more mathematical way, the same
conclusion as above is obtained as follows:
As an alternative to describing the electric field as a superposition of
an infinite number of plane waves one can select electric
and magnetic multipole functions as the basis of choice
\cite{cohen-tannoudji1989,basset1986,vanenk2004}. 
Following Ref.~\cite{basset1986}, it is only the electric-dipole
component that produces a finite electric field at the origin of a
spherical wave.
Therefore, the maximum electric field strength in free space is
produced by an electric-dipole wave.
This field strength is found to be~\cite{basset1986}
\begin{equation}
\label{eq.Emax}
  E_\textrm{max}= \frac{\sqrt{2P}}{\lambda\sqrt{\epsilon_0
      c_0}}\cdot\sqrt{\frac{8\pi}{3}}\quad .
\end{equation}
This is the maximum field strength parallel to the atomic
dipole moment that can be obtained for a given input power $P$ and
wavelength $\lambda$. 
Here, $\epsilon_0$ and $c_0$ are the vacuum permittivity and speed of light,
respectively. 

In practice, the dipole mode is not pure but contains components
from orthogonal electric dipoles and/or magnetic dipoles or higher
order multipoles.
This can be due to a finite solid angle from which the incident light
is focused, deviations of the incident radiation pattern from the
ideal dipole pattern, or a combination of these. 
One can account for such deviations from a pure dipole-mode by
multiplying Eq.~\ref{eq.Emax} with a 
single overlap parameter~\cite{vanenk2004,wang2011}, but from an
experimentalist's perspective it is favourable to distinguish between
different sources of field reduction.
This can be done by introducing two parameters that describe the
experimental geometry~\cite{sondermann2008}. 
The first one is the solid angle covered by the focusing optics
obtained when weighting with the angular intensity pattern of the
atomic dipole:
\begin{equation}
\label{eq.Omega}
\Omega_\mu=\int D_\mu(\vartheta,\varphi)\sin\vartheta d\vartheta
  d\varphi\quad ,
\end{equation}
with the integration performed over the (potentially incomplete) solid angle
set by the focusing geometry.
The symbol $\mu=\pi,\sigma_\pm$ parametrizes a specific dipole
radiation pattern $D_\mu$ given by
$D_\pi(\vartheta)=\sin^2{\vartheta}$ for a linear dipole or
$D_{\sigma_\pm}(\vartheta)=(1+\cos^2{\vartheta})/2$ for a 
circular dipole~\cite{jackson1999}, respectively.
The second parameter is the overlap of the incident field distribution
$\vec{E}_\textrm{inc}(\vartheta,\varphi)$ with the field distribution
of the dipole's radiation pattern $\vec{E}_\mu(\vartheta,\varphi)$,
again obtained by integrating over the focusing solid angle:
\begin{equation}
\label{eq.eta}
\eta=\frac{
\int \vec{E}_\textrm{inc}^\star\cdot\vec{E}_\mu 
\sin{\vartheta}\ d\vartheta d\varphi}
{\sqrt{\int |\vec{E}_\textrm{inc}|^2 \sin\vartheta\ d\vartheta d\varphi
\cdot \int |\vec{E}_\mu|^2 \sin\vartheta\ d\vartheta d\varphi}} 
\quad .
\end{equation}
With these quantities Eq.~\ref{eq.Emax} can be rewritten
as~\cite{sondermann2008}
\begin{equation}
\label{eq.Efoc}
E_\textrm{focus} = \frac{\sqrt{2P}}{\lambda\sqrt{\epsilon_0
      c_0}}\cdot\sqrt{\Omega_\mu}\cdot\eta \quad .
\end{equation}
For sending the ideal dipole pattern ($\eta=1$) and focusing from full
solid angle ($\Omega_\mu=8\pi/3$) Eq.~\ref{eq.Emax} is recovered.

The overlap parameter $\eta$ is very useful in practice. 
In this respect, we should highlight two points:
Since the fields $\vec{E}_\textrm{inc}$ 
and $\vec{E}_\mu$ are complex quantities, it is generally possible to
let $\eta$ also account for distortions of the phase front of the
incident light or aberrations induced by the focusing optics (see also
Ref.~\cite{fischer2014}).
Furthermore, the field distribution of the incident field \emph{after}
transformation into a spherical wave by the focusing optics is usually
not accessible for measurement.
However, $\eta$ can also be calculated in the plane of the entrance
pupil of the focusing optics~\cite{sondermann2008}. 
One just has to calculate how the dipole radiation pattern is
transformed into a plane propagating mode by the focusing device.

Generating the time-reversed version of the latter mode is exactly what
maximizes the interaction Hamiltonian for a given input power and
focusing geometry.\footnote{We consciously neglect effects related to
  a nonzero detuning between the incident field and the atomic
  resonance.} 
Corresponding calculations for a parabolic mirror can be
found e.g. in Refs.~\cite{lindlein2007,sondermann2007,sondermann2008}
and for other optical elements in Ref.~\cite{sondermann2008}. 
The calculation of overlaps of optimized modes generated in an
experiment is treated in Ref.~\cite{golla2012} and in the next section.

We end this section by finding a suitable definition for the 
coupling efficiency in free space.
As outlined in Ref.~\cite{golla2012}, such a definition can be
motivated in analogy to the necessary condition for strong coupling in
cavity quantum-electrodynamics.
The important quantity in this condition is the square of the
single-photon Rabi-frequency of the cavity field.
This frequency is proportional to the atomic dipole moment and the
field obtained from confining the energy of a single photon in the
volume of the cavity mode.
In close analogy, apart from proportionality factors our free-space coupling
efficiency should be given by $|H_I|^2$, which scales with the
intensity of the electric field component polarized parallel to the
atomic dipole moment.
The importance of the mode-matching argument treated above is
highlighted by normalizing $|H_I|^2$ to the ideal
case described by Eq.~\ref{eq.Emax}.
This suggests to define the coupling efficiency $G$
as~\cite{golla2012,leuchs2013o}   
\begin{equation}
\label{eq.G} 
G= \frac{|E_\textrm{focus}|^2}{E_\textrm{max}^2}=
\frac{\Omega_\mu}{8\pi/3}\cdot\eta^2 \quad .
\end{equation}
When reviewing recent experiments in free space, we will highlight the
role of the coupling efficiency in more detail.

\section{Dipole-mode generation with a parabolic mirror}

\subsection{Finding the optimum field mode}

From the discussion in the previous section it has become obvious that
one should focus from (almost) the entire solid angle.
One suitable setup immediately coming to mind is the setup of the
so-called $4\pi$-microscopy~\cite{hell1992} which consists of two high
NA lenses with coinciding focal points.
Although some of the experiments performed in recent years in
principle provide the possibility for focusing with two
lenses~\cite{tey2008-np,piro2011,aljunid2013} and thus doubling the solid angle
coverage, such an experiment has not been reported.
This might be due to technical issues related to the necessity of
achieving constructive interference between the waves focused by the
two objectives. 
But this might also be due to the large band of solid angles located
symmetrically around the optical axis which are not covered by the
highest numerical aperture (NA) of a commercial objective in vacuum,
i.e. NA=0.95.  
In the best case, the missing 0.05 to achieve NA=1 corresponds to
losses of 24\% for the weighted solid angle for a $4\pi$-microscopy
setup. 
Moreover, in most experiments, especially the ones on trapped ions,
the used NA is considerably smaller.

\begin{figure}
\includegraphics{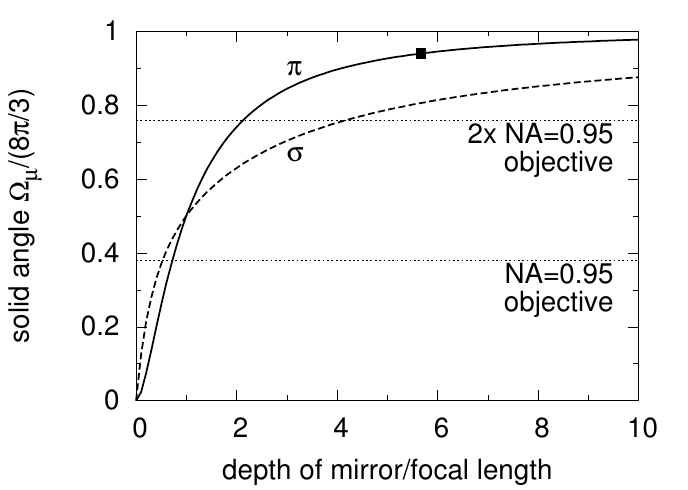}
\caption{\label{fig.SA}Weighted solid angle $\Omega_\mu$ covered by a
  parabolic mirror.
The solid (dashed) line denotes the case of a linear (circular) dipole
transition with the quantization axis parallel to the optical axis of
the mirror.
The symbol marks the geometry used in our experiments.
The dotted lines display $\Omega_\mu$ for one and two NA=0.95 lenses
and a linear dipole transition with the quantization axis
perpendicular to the optical axis (best case, worse for other
orientations).} 
\end{figure}

Another possibility for focusing from essentially full
solid angle is using a parabolic mirror that has a depth much
larger than the focal
length~\cite{lindlein2007,sondermann2007,gardiner1986,bokor2008}.
As is obvious from Fig.~\ref{fig.SA}, when the ratio of the depth $h$ of
the parabola and the focal length $f$ surpasses the limit $h/f\approx
2.12$ a parabolic mirror setup outperforms $4\pi$-microscopy setups.
It is also evident from the figure that the configuration of a
$\pi$-transition with the quantization axis parallel to the mirror's
optical axis yields the largest values of $\Omega_\mu$ for finite
sized parabolic mirrors covering more than half of the solid
angle.\footnote{The half solid angle case corresponds to $h/f=1$.} 
Therefore, we have chosen to use this configuration in our
experiments~\cite{leuchs2008,golla2012,maiwald2012,fischer2014}.

\begin{figure*}
\centerline{\includegraphics{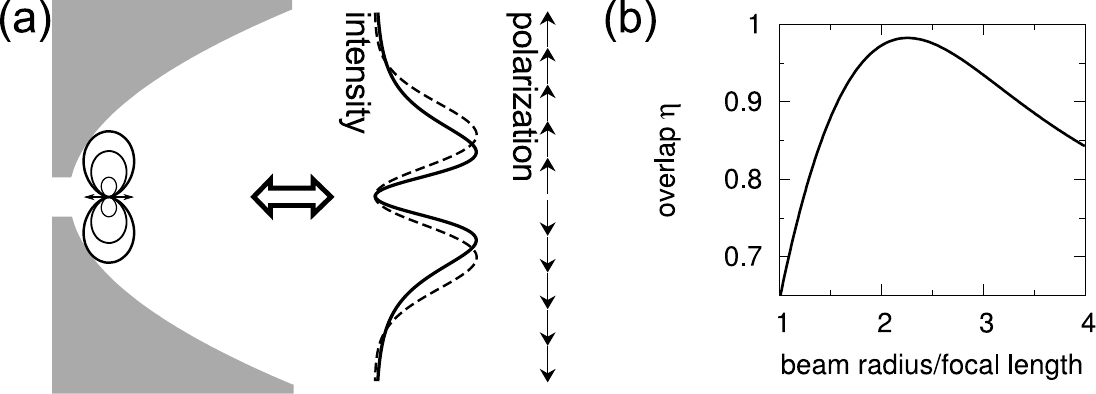}}
\caption{\label{fig.parabola} 
(a) Basic layout of a setup optimizing light-matter coupling in free
  space. A single atom is located in the focus of a parabolic mirror
  with depth much larger than its focal length. A beam with the
  intensity distribution given by the square of Eq.~\ref{eq.Elin}
  (solid line) and with radial polarization (arrows) is focused by the
  parabolic mirror. For comparison, the intensity distribution of the
  doughnut mode with largest overlap $\eta$ is also displayed (dashed
  line).   
(b) Overlap of a radially polarized doughnut mode with the ideal dipole
  distribution as a function of the doughnut mode's beam radius for the
  geometry used in our experiment ($h/f=5.67$ and
  $\Omega_\mu=0.94\cdot8\pi/3$).
}
\end{figure*}

Next, we derive the field distribution to be sent onto the
parabolic mirror for optimized coupling to an atom in the above
configuration. 
Following Refs.~\cite{lindlein2007,sondermann2008}, one has to trace
the emission pattern of the atomic dipole-radiation towards the
parabolic surface, relating emission angles $\vartheta$ to distances
$r$ from the optical axis by $\vartheta=2\arctan(r/(2f))$. 
Furthermore, one has to account for the proper transformation of power
emitted per solid angle into power per surface area.
This finally yields the amplitude distribution of a linear dipole
collimated by the parabolic mirror:
\begin{equation}
\label{eq.Elin}
E_\pi(r)= E_0\cdot\frac{r}{(r^2/(4f)^2+1)^2}\quad .
\end{equation}
Likewise, one finds the corresponding polarization pattern. 
For our configuration, the field in the exit pupil of the parabolic
mirror is radially polarized.
These findings are illustrated in Fig.~\ref{fig.parabola}.

The generation of the mode defined by Eq.~\ref{eq.Elin} might be
rather intricate~\cite{lindlein2007}.
Fortunately, there exists a mode that can have large overlaps with the
ideal mode~\cite{quabis2000,lindlein2007,sondermann2008} and which is
routinely generated in experiments: a radially polarized
mode with the amplitude distribution of a Laguerre-Gaussian beam of
zeroth radial and first azimuthal order.
In the following, we will call this mode shortly \lq doughnut
mode\rq\ for simplicity, although the latter term is used in
literature for a plethora of modes with 
ring shaped amplitude pattern. 
The amplitude distribution of a doughnut mode can be written as 
\begin{equation}
\label{eq.Edonut}
E(r)= E_0\cdot r\cdot \mathrm{e}^{-\frac{r^2}{w^2}}
\end{equation}
with the beam radius $w$.
For any given parabolic mirror, the overlap parameter $\eta$ is
maximized by tuning $w$~\cite{sondermann2008}.\footnote{This is of
  course also true for other focusing optics or 
  dipole configurations and correspondingly other suitable 'standard'
  field modes as e.g. a fundamental Gaussian mode, see
  Ref.~\cite{sondermann2008} for some examples.}
For our geometry ($h/f=5.67$ and $\Omega_\mu=0.94\cdot8\pi/3$) the
overlap is maximized by $w=2.26f$ to be $\eta=98.2$\% (see
Fig.~\ref{fig.parabola}b).
The intensity profile of this optimum doughnut mode is displayed as a
dashed line in Fig.~\ref{fig.parabola}a.
Using the above values of $\Omega_\mu$ and $\eta$, we could in
principle achieve overall coupling efficiencies of $G\approx0.9$ with
our setup.

However, one has to keep in mind that expanding the incident beam
towards a size which maximizes $\eta$ might result in clipping
considerable parts of the beam and hence in losses, which can be
considered as a \lq no-go\rq\ when focusing single photons onto an atom.
This is a considerable effect in setups with small to medium sized
numerical apertures, e.g. the NA=0.4 objectives used in
Refs.~\cite{slodicka2010,piro2011}. 
Hence it is required to maximize the product of the power transmitted
through the aperture and $\eta^2$ for a given entrance pupil.
However, for the parabolic mirrors used in our experiments the
relative power loss for the doughnut mode with maximum $\eta$ is on the
order of $10^{-3}$.
We therefore can safely neglect such effects.

On the other side, when using doughnut modes for focusing with
parabolic mirrors it is not too meaningful to use parabolic mirrors
with even larger solid angle coverage than in our setup: The gain of
solid angle $\Omega_\mu$ is compensated by a reduction of the maximum
value attainable for $\eta$, which results in a saturation of the
coupling efficiency at $G\approx0.92$~\cite{sondermann2008}.
Of course, the latter value depends on the choice of the focusing
geometry, the dipole configuration, and the optical mode used to
approximate the ideal dipole radiation.

\subsection{Generation and characterization of field modes tailored
  for efficient free-space coupling}

We now turn towards the experimental generation and characterization
of the optimum doughnut mode for focusing onto an atom.
There are various methods for generating radially polarized doughnut
modes~\cite{bomzon2002,ghadyani2011,stalder1996,tidwell1990,
  tidwell1993,oron2000,maurer2007,lin2013}.
We have chosen to generate the doughnut mode by means of a segmented
half-wave plate~\cite{dorn2003,quabis2005}.
This is mainly motivated by the fact that this technique is rather
robust and especially suitable for wavelengths in the ultraviolet
spectral range, which we are
targeting~\cite{sondermann2007,leuchs2008,golla2012,maiwald2012}.
For details about the actually used polarization converter and the
optical setup we refer to Ref.~\cite{golla2012}.

\begin{figure*}
\centerline{\includegraphics{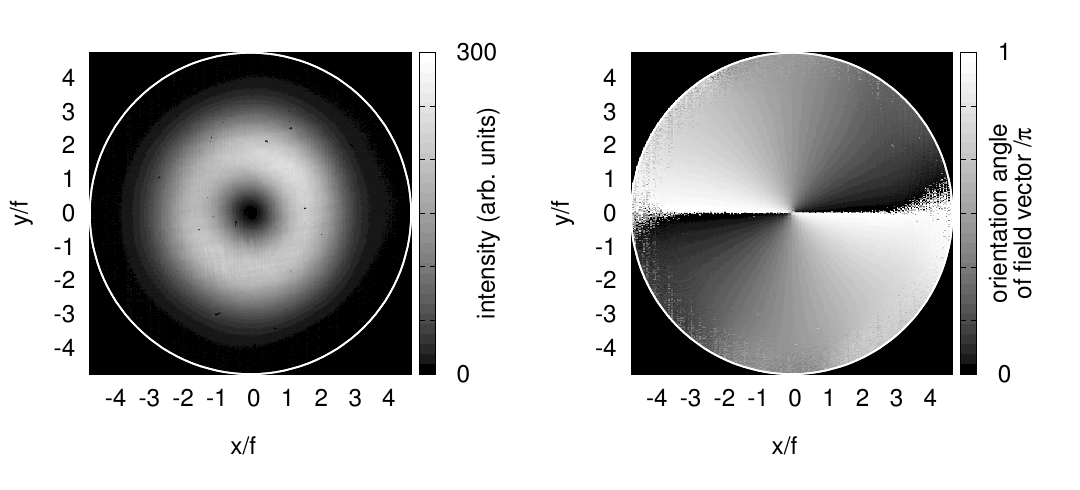}}
\caption{\label{fig.donut}
Example of a radially polarized doughnut mode generated at a
wavelength of 370~nm and optimized to drive the
S$_{1/2}$,m$_j=\pm1/2\rightarrow$P$_{1/2}$,m$_j=\pm1/2$ transition
of a singly ionized ytterbium ion. 
The intensity distribution (left) and the orientation angle of the
local polarization ellipse of the field vector (right) are
reconstructed from a spatially resolved measurement of the Stokes
parameters.
The circle indicates the entrance pupil of the
parabolic mirror.}
\end{figure*}

One example of a generated mode is given in Fig.~\ref{fig.donut}.
The local intensity as well as the local polarization ellipse of the
electric-field vector are obtained performing a spatially resolved
measurement of the Stokes parameters (the ellipticity angle is omitted
in the figure, cf. Ref.~\cite{golla2012} for details).
Taking the square root of the local intensity and processing the
polarization angles one can reconstruct the local electric-field
vector.
With the obtained field vectors one can compute the overlap $\eta$
with the ideal field distribution given by Eq.~\ref{eq.Elin}.
We routinely achieve values of $\eta=0.98$~\cite{golla2012} which is
practically the maximum value attainable in our setup.

So far we did not discuss the properties of the phase front of the
incident field.
Under ideal conditions, a spherical wave emerging from the focus of
the parabolic mirror leaves the mirror's aperture with a flat phase
front.
Likewise the phase front of the incident field that impinges onto the
mirror and is focused onto the atom should be flat.
Unfortunately, current mirror manufacturing capabilities are not
precise enough to guarantee the diffraction limited focusing of an
incident beam that is free of aberrations.
The mirrors we use in our experiments typically deviate from
the parabolic shape by $\pm$150\,nm~\cite{leuchs2008}, which is on the
order of a wavelength of the involved atomic transitions.
We determine such deviations by an interferometric setup that is
adapted to our purpose~\cite{leuchs2008}.
Other methods involving e.g. profilometers are not applicable due to
the tight geometry of the used mirrors -- the focal length is 2.1\,mm
and the aperture radius is 10\,mm.

In principle the deviations of the parabolic mirror can be compensated for
by using phase plates which imprint the conjugate of the mirror's
deviations onto the incident beam.
Proof-of-principle experiments at a wavelength of 633\,nm yielded
aberration compensations enabling a \emph{Strehl ratio}\footnote{
The Strehl ratio defines the maximum intensity in the focal region for
a focusing system exhibiting aberrations as a fraction of the
intensity obtained without aberrations~\cite{born-wolf1991}. Whereas
in the latter reference a plane wave is considered, we apply this
figure of merit for the case of a doughnut mode.}
of 99\% for a focused doughnut mode at 633\,nm~\cite{golla2012}.

\section{Overview of experiments on photon-atom coupling in free-space}

In this section we discuss various kinds of experiments on 
photon-atom coupling in free space performed
in recent years. 
They all have in common that the key to a successful observation of
the phenomena under investigation lies in mode-matching the
incident field to the dipolar radiation pattern of the addressed
atomic transition. 
We will highlight for each type of experiment how it could benefit
from full solid angle focusing or the use of parabolic mirrors,
respectively.

\subsection{Shifting the phase of a coherent beam}

For the experiments described here -- and also for the ones on
attenuating a faint beam -- the typical experimental setup is as
follows~\cite{aljunid2009,pototschnig2011,hetet2013}:
A continuous laser beam is focused onto the atom.
Depending on the detuning $\Delta$ of the incident light with respect
to the atomic resonance a certain amount of light is scattered by the
atom with a detuning dependent phase shift\footnote{
The phase shift of the light \emph{scattered} by an atom, i.e. the
response of a driven harmonic oscillator, has been measured
recently~\cite{jechow2013}.}. 
The light passing the atom and the scattered light are both collected
by some optical element, e.g. a second lens of the same numerical
aperture as the focusing optics.
Upon diverging from the atom, the incident light suffers a phase shift
of 90$^\circ$ due to the Gouy
effect~\cite{zumofen2008,tey2009,tyc2012}.
For simplicity, the latter contribution is artificially attributed to
the phase of the scattered light in most literature.
The phase that is then usually measured is the phase of the
superposition of the scattered light and the incident light.

\begin{figure*}
\centerline{\includegraphics{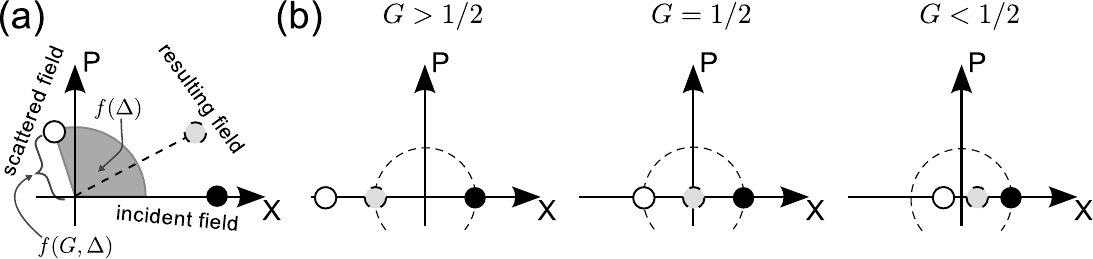}}
\caption{\label{fig.phasespace}
Illustration of phase shift experiments in phase-space diagrams.
(a) Depending on the detuning $\Delta$ and the coupling efficiency $G$
the field scattered by the atom (open circle) has a certain amplitude
and phase. 
The superposition of the incident field (filled circle) and the
coherently scattered field (open circle) yields the resulting field
(dashed circle).
(b) Illustration of the influence of $G$ for the special case
$\Delta=0$. 
The phase of the resulting field can either be equal to $0$ or $\pi$.
The dashed line marks the amplitude of the incident coherent state.}
\end{figure*}

This scenario is illustrated in Fig.~\ref{fig.phasespace}a in a
phase-space picture. 
The phase angle of the scattered field is solely determined by the
detuning $\Delta$, whereas the amplitude of the scattered field is
determined by $\Delta$ and especially $G$.
The latter statement is obvious from the fact that the power of the
scattered light in terms of power of the incident light is given by
$4G/(1+4\Delta^2/\Gamma^2)$~\cite{leuchs2013o,sondermann2013p} with
$\Gamma$ being the rate of spontaneous emission from the excited
atomic state.
This so-called \emph{scattering ratio}~\cite{zumofen2008,tey2009} can
be as large as four for a perfect dipole-wave incident from full solid
angle and on resonance with the atomic transition.\footnote{Values
  for the scattering ratio exceeding unity might seem unphysical 
at first sight because one might suspect a violation of energy
conservation. Here we argue that this is not the case. Energy is
always associated with the total
field not with individual interfering components. One might write a
propagating field as the sum of two fields $E_1$ and $E_2$, one
180$^\circ$ out of phase with respect to the other. The total energy
has of course a well defined value. But the individual fields
$E_{1,2}$ are not well defined. 
You can choose a field with a larger $E_1$ as long as the
amplitude of field $E_2$ is also increased such that the sum is as
before. 

In the specific problem of elastic scattering of a beam resonant with
the atomic transition it can even be required that the scattered
power has to be larger than the incident one.
Due to the 180$^\circ$ phase shift between the scattered and the
transmitted incident light there would otherwise be a net loss of
energy under efficient-coupling conditions, since in elastic
scattering no energy is transferred onto the atom.
} 
From the fact that the scattering ratio scales with the coupling
efficiency $G$ one can draw the following conclusion:
No matter how large the detuning is, the influence of the scattered
light on the final superposition with the incident light grows with
increasing $G$ and is maximized for $G=1$,
cf. Fig.\,\ref{fig.phasespace} and the discussion following
Eq.\,\ref{eq.phases}.
In other words, the phase obtained for the superposition field \lq is
drawn\rq\ towards the phase of the scattered field with increasing $G$.

In the discussion so far we have not included effects related to a
finite amount of population in the atom's excited state.
The excited-state population can be quantified by $S/(2+2S)$ using the
\emph{saturation parameter} $S$, which is proportional to
$G$ and depends on detuning, see also Eq.\,\ref{eq.saturation}.
This parameter influences the phase shift in two ways:
For increasing saturation parameter the scattering ratio decreases.
Furthermore, the scattered light contains an increasing fraction of
frequency components that are not coherent with the incident light and
hence cannot contribute to the superposition with the incident field.
A corresponding experiment can be found in Ref.~\cite{pototschnig2011}.
We summarize the above discussion, assuming negligible saturation, in
writing the resulting phase shift, i.e. the phase of the superposition
of scattered and incident light in comparison to the case of no atom
being present, as~\cite{leuchs2013o,sondermann2013p} 
\begin{equation}
\label{eq.phases}
\phi=\arg\left(1 + 4\frac{\Delta^2}{\Gamma^2}-2G
-i\cdot4G\frac{\Delta}{\Gamma} \right) 
\end{equation}

It is apparent from the above equation that the maximum achievable
phase shift is $\phi=\pi$ for $\Delta=0$ (resonant excitation) and $G>1/2$.
This phase arises from a $\pi/2$ phase for the coherently scattered
resonant light and the Gouy phase.
However, for $G<1/2$ (less than half solid angle and/or too low field
overlap) the amount of scattered light is too small to counteract the
incident field. 
For $G>1/2$ (focusing from more than half solid angle with
sufficiently mode-matched incident light), the amount of scattered
light is large enough to flip the phase of the superposition
with the incident field.
A corresponding illustration is given in Fig.~\ref{fig.phasespace}b.

Although it seems that it is enough to have \lq $G$ just slightly larger
than $1/2$\rq, there are good reasons for increasing $G$ towards unity.
The closer $G$ is to $1/2$ the smaller is the detuning
interval in which $\phi$ increases from $\pi/2$ towards $\pi$, making
an experimental observation of a $\pi$ phase shift increasingly
difficult (cf. Ref.~\cite{sondermann2013p}).

The maximum phase shift observed so far in an experiment in
free space was about 3$^\circ$~\cite{pototschnig2011} using a setup
with $G<1/2$. 

\subsection{Extinction of a weak coherent beam}

An experiment closely related to imprinting a phase shift onto a
coherent beam is attenuating such a beam with a single
atom\cite{vamivakas2007,wrigge2008,tey2008-np,tey2009, 
slodicka2010,aljunid2011,rezus2012}.
In such experiments the incident beam is typically focused and
recollimated with two lenses of equal numerical aperture.
In the absence of the atom all light is transmitted through the
two-lens setup.
If an atom resides in the focus of the incident beam, the interference
between incident and scattered light results in a decrease of the
number of transmitted photons.
The exact strength of this extinction depends again on the
scattering ratio~\cite{zumofen2008,tey2009}:
For focusing from half solid angle with a properly mode matched wave
resulting in $G=1/2$, the scattering ratio reaches two, with the
consequence that the incident light and the light scattered into the
solid angle cone of the second lens interfere destructively.
In other words, a single atom can perfectly reflect a resonant
coherent beam~\cite{zumofen2008,tey2009,kochan1994}.
The partial reflection of an incident beam was measured in
Ref.~\cite{aljunid2011}.
In Ref.~\cite{hetet2011} the back scattering from a single ion was
exploited for utilizing a single ion as a mirror of an optical
resonator (see also the chapter by
\href{http://arxiv.org/abs/1502.04302}{Slodicka et al}).
The smallest transmission observed so far for a single quantum emitter
in free space was about 78\%~\cite{wrigge2008}.

The transmitted power fraction $T$ of a resonant incident beam
inducing negligible saturation of the atom can be derived
straightforwardly, again assuming the same numerical aperture for
focusing and light collection:
The scattering ratio is given by $4G$, i.e. the light scattered into
the solid angle outside the collection optics is given by
$4G\cdot[1-\Omega_\mu/(8\pi/3)]$.
This yields
\begin{equation}
\label{eq.transmission}
T= 1 - 4G\cdot\left(1-\frac{\Omega_\mu}{8\pi/3} \right)\quad .
\end{equation}
Obviously, for $\Omega_\mu=8\pi/3$ one gets full transmission. 
This solution is however trivial, since for a full-solid-angle optics no
photons are lost due to scattering.

\begin{figure}
\includegraphics{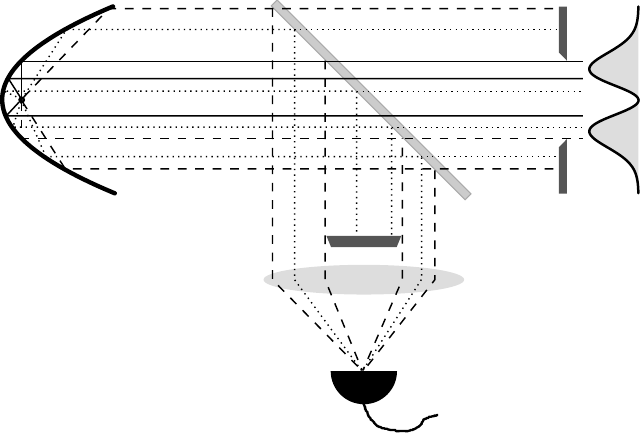}
\caption{\label{fig.extinction}
Layout of an extinction experiment using a parabolic mirror.
The incident light (solid lines) is limited to half solid angle by an
aperture with diameter $4f$.
Light passing the atom (dashed lines) and the scattered light
(dotted lines) are picked off by a beam splitter. 
Light scattered \lq backwards\rq\ is blocked by an aperture stop of diameter
$4f$.}
\end{figure}

More interesting is the case of $T=0$, which can be realized with a
full-solid-angle parabolic mirror.
If the incident light is transversely limited to exactly half of the
solid angle (the portion of the parabola defined by $r\le2f$) one has
$\Omega_\mu=4\pi/3$. 
If furthermore $\eta=1$ one obtains $G=1/2$ and hence $T=0$
according to Eq.~\ref{eq.transmission}.
In such a setup, the part of the mirror with $r>2f$ acts as a second
\lq objective\rq\ collimating the transmitted light.  
A possible layout of such an experiment is sketched in
Fig.~\ref{fig.extinction}.

\subsection{Absorption of single photons}

In the experiments discussed so far the light focused onto the atom
was of sufficiently low intensity to avoid a non-negligible excitation
of the atom, i.e., the number of photons per excited-state lifetime
was much smaller than one.
In contrast to this, in the experiments reported in
Refs.~\cite{piro2011,aljunid2013} the explicit aim was to excite
the atom with single photons.
In Ref.~\cite{piro2011} the absorption of a heralded and spectrally
filtered single photon by a single ion is reported (see chapter by
\href{http://arxiv.org/abs/1502.04349}{Piro and Eschner}).
The reported absorption efficiency is on the order of 0.03\%.
In Ref.~\cite{aljunid2013} weak coherent states have been focused onto
a single atom and an absorption efficiency of about 4\% is reported
for pulses containing two photons on average.

As outlined in Ref.~\cite{golla2012} the probability for the
absorption of a photon by a two-level atom is given by 
\begin{equation}
\label{eq.absorb}
P_\textrm{a}= G\cdot\eta_t^2\quad .
\end{equation}
The parameter $\eta_t$ describes the temporal overlap of the field
envelope of the incident single photon with the field envelope that
maximizes absorption.
The latter is given by an increasing exponential pulse, as described in
detail in the next section.
According to Eq.~\ref{eq.absorb} the upper bound for the absorption of
single photons by atoms in our parabolic mirror setup is given by
$G=0.94$ when sending the ideal radiation pattern and $G=0.9$ when
sending an optimized radially-polarized doughnut mode.

We finish this section highlighting a link between the elastic
scattering of a monochromatic wave discussed before and the absorption of
photons, see Ref.~\cite{sondermann2013s}.
For this purpose we treat the interaction of the atom with all
spectral components of the focused pulse as an elastic scattering
problem.
We compute the phase and amplitude of the superposition of
the incident field and the coherently scattered field for each spectral
component of the incident pulse.
This yields the spectrum and hence the temporal evolution of
the scattered pulse. 
Performing such a calculation for an exponentially increasing pulse with
a time constant matching the excited state's lifetime yields an
outgoing pulse envelope containing exponentially increasing and
decreasing components.\footnote{The decreasing components should of
  course be observable for any incident pulse with non-zero temporal
  overlap $\eta_t$.}
The fraction of the exponentially decreasing components is given by
$G$, quantifying the amount of absorbed and spontaneously re-emitted
photons~\cite{sondermann2013s}. 
For $G=1$ full absorption is obtained, as one would expect from
time-reversal symmetry arguments discussed below. 
But this finding should be taken with care, since time-reversal
symmetry arguments demand the use of a single-photon Fock state.
Nevertheless, the above treatment should be valid for weak coherent
state pulses as used in Ref.~\cite{aljunid2013} or the ones created in
Ref.~\cite{golla2012}.

\section{Absorbing a single photon: temporal mode shaping}

\subsection{Choosing the right mode}

\begin{figure}
\includegraphics{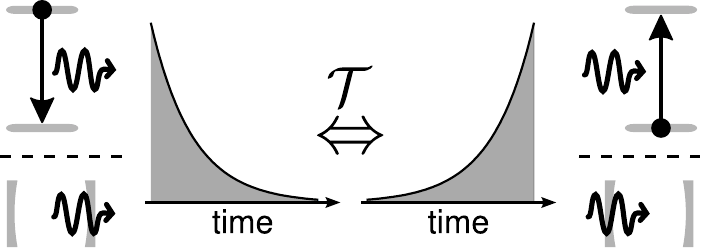}
\caption{\label{fig.reversal}
Illustration of the time-reversal arguments for coupling to a single
atom or into an empty optical resonator as treated in
Sec.\,\ref{sec.resonator}.}
\end{figure}

As mentioned above, the temporal envelope of the focused electric field
is of importance when exciting a single atom with a single photon.
This can be motivated heuristically as follows (see also
Refs.~\cite{quabis2000,sondermann2007,leuchs2012}):
The atom and the continuum of all free-space field modes form a closed
system. 
For such a system, the Schr\"odinger equation is invariant under time
reversal.
Assuming the atom in its excited state and the electro-magnetic field
in the vacuum state as the initial condition, a photon will be emitted
spontaneously over the course of time.
As is well known, a spontaneously emitted photon has an exponentially
decaying field envelope with a decay constant equal to the
excited-state lifetime. 
Applying the time reversal operation -- we just assume we could
do so in practice -- would result in an exponentially increasing
photon travelling towards the atom, promoting the atom to the excited
state as outlined in Fig.~\ref{fig.reversal}.

Applying the time-reversal operation in practice, which
implies phase-conjugating a single photon, is carefully speaking
subject to technical difficulties -- and also the evolution in the
atom's motional degrees of freedom has to be handled~\cite{leuchs2012}.
Moreover, the phase-conjugation process induces excess quantum
noise~\cite{yamamoto1986,gaeta1988}.
It is therefore the idea to shape a photon to resemble a perfect copy
of a time-reversed spontaneously emitted photon.\footnote{It can be
  shown in a fully quantum-mechanical calculation that an
  exponentially rising pulse with proper time-constant indeed 
  leads to full excitation of the atom~\cite{stobinska2009}.}
This implies the spatial mode-matching arguments raised in
Sec.~\ref{sec.dipole} but also requires \lq temporal mode-matching\rq, as
expressed by the temporal overlap-parameter $\eta_t$ in
Eq.~\ref{eq.absorb}.

In practice, the assumption of working with two-level atoms is usually
not justified and there is more than one possible decay channel from
the excited state.
Hence, spontaneous emission would result in a superposition
of the decays via all these channels, entangling the emitted photon
and the remaining atom.
The reverse process, absorption of a photon, would ideally start from
such an entangled state.
This is extremely challenging if not self forbidding.
We therefore seek an atomic species providing as small as possible
branching ratios in the excited state's decay as well as a
$\pi$-transition.
Any even numbered isotope of doubly ionized ytterbium (YbIII) is such
a species.
It is also a good choice from a practical point of view,
since singly-ionized ytterbium (YbII) is a well-established ion used
in many quantum-optics experiments and for atomic clocks.
YbII also offers a strong $\pi$-transition (see Fig.~\ref{fig.Yb}),
serving as our test-system for all techniques enabling efficient
free-space coupling.\footnote{YbIII has been created by electron-impact
  ionization from a cloud of trapped YbII ions~\cite{schauer2010}. We
  recently accomplished the controlled photo-ionization
  from YbII to YbIII.}

\subsection{Generation of exponentially increasing pulses}

The fact that most ions have to be addressed by ultraviolet light
makes the generation of single-photon Fock states with proper temporal
envelope a cumbersome task (see Ref.~\cite{golla2012} for citations on
  shaped photons with more \lq accessible\rq\ wavelengths).
Nevertheless, the use of whispering-gallery-mode
resonators~\cite{foertsch2013} made of appropriately chosen materials
might facilitate shaped single photons at UV wavelengths.

For the time being and as a preliminary test, we resort to using
suitably shaped weak coherent states~\cite{golla2012}.
We generate these states by modulating a continuous laser beam with an
acousto-optic modulator (AOM).\footnote{A similar technique employing
  electro-optic modulation is reported in Ref.~\cite{dao2012}.
In contrast to this, other authors directly modulate the waveform of a
single photon upon receiving a trigger signal from a heralding photon,
see Ref.\,\cite{kolchin2008} and the chapter by 
\href{http://arxiv.org/abs/1503.04339}{Chuu and Du}.
Last but not least, in a recent experiment single-photon Fock states
with increasing exponential envelope have been achieved by
manipulating a heralding photon with an asymmetric cavity prior to
detection\,\cite{srivathsan2014}. Both photons originated from a
cascaded decay in a cold atomic ensemble.} 
The AOM is driven with a voltage signal generating exponentially
rising pulses in the first diffraction order. The voltage signal is
proportional to
$\arcsin[\exp(t/(2\tau))]\cdot\sin(\omega_\textrm{RF}t)$
,\footnote{This expression arises from the fact that the power
  scattered into the first diffraction order of the AOM is proportional
  to the sine of the acoustic power establishing the diffraction grating.}
where
$\tau=1/\Gamma$ is the excited-state lifetime and $\omega_\textrm{RF}$
the frequency with which the AOM is driven.
The obtained pulses are attenuated towards mean photon numbers about
0.1 and characterized by statistics on the times of photon detection
events obtained from a photo-multiplier tube,
cf. Ref.\,\cite{golla2012} for details.

\begin{figure}
\includegraphics{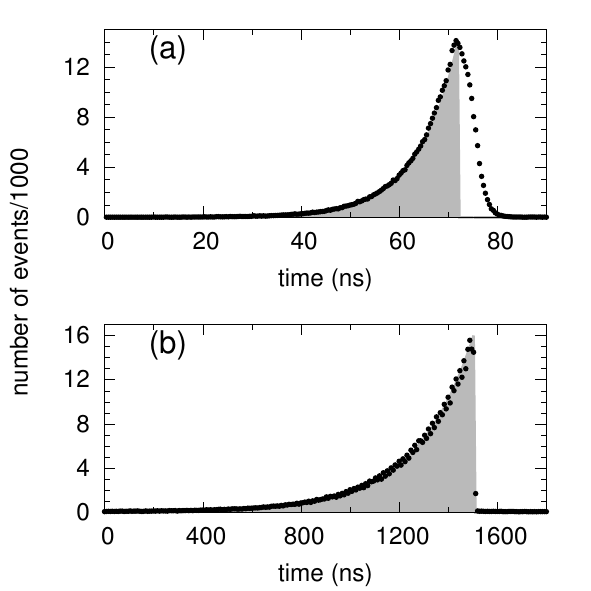}
\caption{\label{fig.pulses}
Weak coherent-state pulses generated to drive an atomic transition
with 8\,ns excited-state lifetime (YbII $^2P_{1/2}\rightarrow\,^2S_{1/2}$, a)
and 230\,ns excited-state lifetime (YbIII $^3P_1\rightarrow\,^1S_0$,
b).
The average photon number per pulse is about
0.1. Histogram data of photon detection times are denoted by 
circles. Each histogram contains more than $4\cdot10^5$ events. The
shaded area depicts the optimum pulse shape.}
\end{figure}

From the obtained histograms (see Fig.~\ref{fig.pulses}) we
reconstruct the field amplitude's envelope $E_\textrm{inc}(t)$ by
taking the square root of the number of events per time bin.
We then compute the overlap of $E_\textrm{inc}(t)$ with the ideal
envelope $E_\textrm{ideal}(t)=e^{\frac{\Gamma}{2}t}\cdot\theta(-t)$
obtained from time-reversal symmetry arguments:
\begin{equation}
\label{eq.tempovl}
\eta_t=\frac{\int_{-\infty}^\infty E_\textrm{inc}(t) \cdot
   E_\textrm{ideal}(t) dt} 
{\sqrt{\int_{-\infty}^\infty |E_\textrm{inc}(t)|^2 dt / \Gamma}}
\quad .
\end{equation}
For the data shown in Fig.~\ref{fig.pulses} we obtain 
$\eta_t=0.96$ for the YbII transition and $\eta_t=0.99$ for the YbIII
transition.
The slightly lower overlap in the case of the YbII transition is due
to the fact that the drop-off at the end of the pulse, which amounts
to 5\,ns enforced by the finite decay time of the acoustic grating
inside the AOM, is on the order of the lifetime of 8\,ns.
Hence, the drop of intensity at the end of the pulse is not as
step-like as required.
Nevertheless, both pulse shapes envision absorption probabilities
close to $G$, neglecting the non-ideal level structure of
YbII~\cite{golla2012}.

\subsection{An analogous experiment: coupling to a resonator}
\label{sec.resonator}

An experiment conceptually analogous to absorbing a single photon --
but with significantly reduced technical complexity -- is the coupling
of pulses into an empty optical resonator~\cite{heugel2010}. 
Imagine a finite amount of energy being stored inside the resonator
at a certain point in time.
This electro-magnetic field will leave the resonator via exponential
decay, where the time constant of the decay is given by the decay time
of the intra-cavity field.
If the resonator consists of only two mirrors with one of them having unit
reflectivity, the field leaves the cavity through the mirror with
sub-unit reflectivity.
Now time-reversal arguments tell us that the energy of a properly
shaped, exponentially increasing pulse sent towards the cavity 
is stored completely inside the cavity. 
In other words, no light is reflected from the cavity as long as the
incident pulse continues to grow~\cite{heugel2010}.

\begin{figure}
\includegraphics{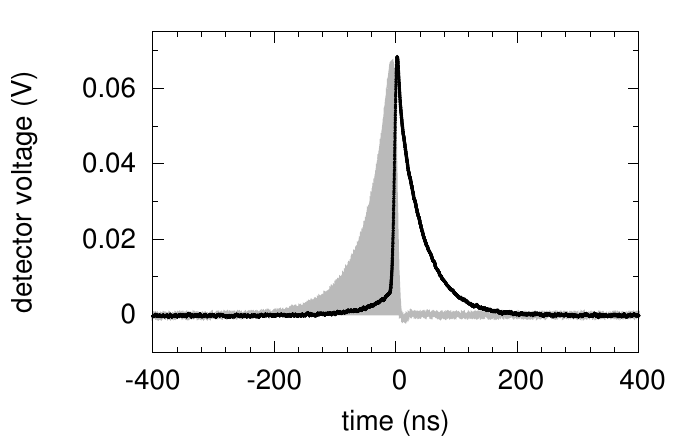}
\caption{\label{fig.cavity}
Response of an empty optical resonator to a pulse with increasing
exponential envelope and a time constant matching the cavity's
decay-time.
Black line: reflected signal for resonant excitation.
Shaded area: reflected signal for large detuning from the cavity
resonance, being in good approximation identical with the incident
pulse.} 
\end{figure}

Recently, we have implemented a corresponding
experiment~\cite{bader2013}, shaping the pulses in the same way as
described above in the generation of the weak coherent-state pulses.
The non-confocal resonator, which was used in our experiment,
consisted of a mirror with a reflectivity of $R_1=97.96$\% and
$R_2=99.94$\% and had an intensity decay time of 39\,ns.
In the experiment, incident light was spatially mode matched to
the resonator mode, sent onto the resonator via the mirror with
reflectivity $R_1$ and tuned into resonance.
The temporal envelope of the reflected pulse was monitored with a
photo diode.
The outcome of such an experiment is presented in
Fig.~\ref{fig.cavity}.

The finite amount of light leakage through the second mirror,
preventing measurements of the cavity transmission with sufficient
bandwidth, as well as non-perfect spatial mode-matching necessitate a 
somewhat more involved analysis~\cite{bader2013}. 
Despite the imperfections, we achieved 88\% energy storage
efficiency for the generated pulses and even 94\% efficiency for the
spatially mode-matched fraction.\footnote{Similar efficiencies where
  obtained recently in experiments for microwave
  pulses~\cite{palomaki2013,wenner2014}.} 
Deviations from unity are due to a non-unit temporal field overlap
$\eta_t=0.986$ and leakage through the second mirror.
In analogy to the single-atom experiment this leakage corresponds to a
solid-angle coverage of 97\%.
The achieved large efficiencies hint at the power of
time-reversal-symmetry based arguments when optimizing the coupling of
light and matter.
As a further example, true single-photon Fock states with increasing
exponential temporal envelope have been efficiently coupled to a cavity
recently, c.f. Ref.~\cite{liu2014}. 

\section{Trapping ions in parabolic mirrors}

\subsection{Parabolic mirror ion trap}

\begin{figure}
\includegraphics{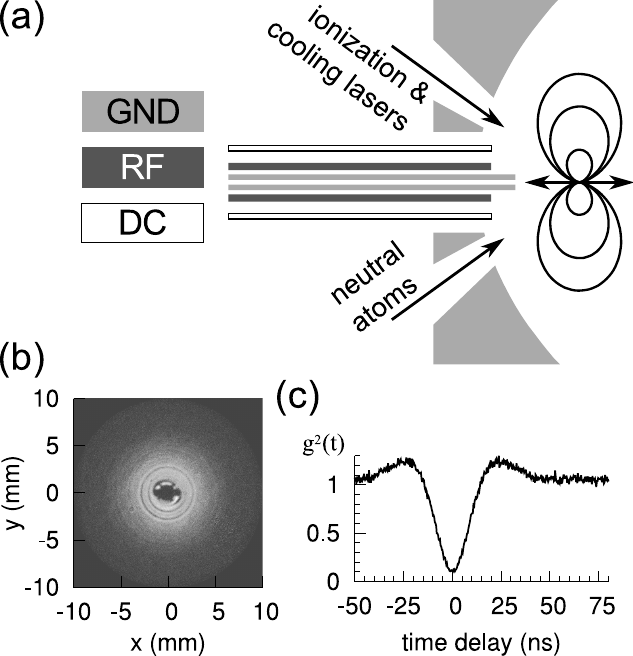}
\caption{\label{fig.trap}
(a) Schematics of the parabolic-mirror ion-trap, consisting of
two concentric tubular electrodes for the radio-frequency (RF)
signal and ground (GND) potential. The aluminium
parabolic mirror is also grounded. Rod electrodes (DC) enable
micro-motion compensation.
(b)~A single YbII ion's fluorescence imaged in the
aperture plane of the parabolic mirror. White colour indicates
large intensity.
(c) Second-order intensity correlation function (anti-bunching)
obtained from approx. $8\cdot10^5$ fluorescence counts in 60\,s.
}

\end{figure}

\begin{figure}
\includegraphics{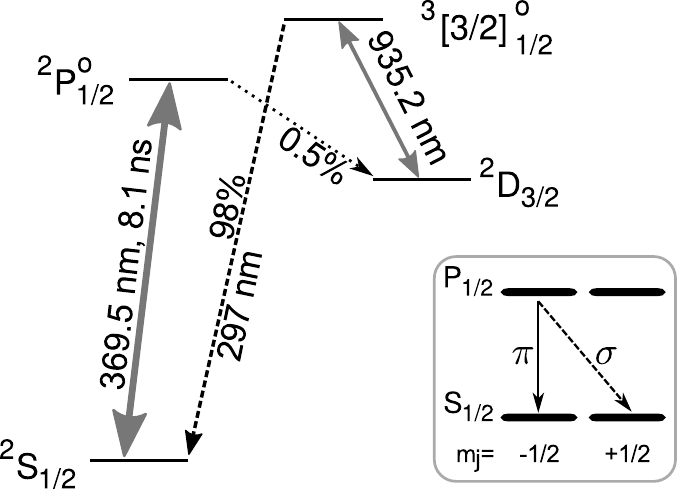}
\caption{\label{fig.Yb}
Energy levels of YbII most relevant for our experiments.
The inset details the levels addressed in our photon-atom coupling
experiments.}
\end{figure}

So far we did not discuss a most important issue of photon-atom coupling with
parabolic mirrors: How do we place the atom at the focus of the
parabolic mirror?
As already mentioned above, we have chosen to work with atomic ions.
This demands for an ion trap potential that can be located precisely
enough to place the ions within the tight focal spot of the incident
beam.
Simultaneously, the ion trap geometry should maintain the large
solid-angle optical access to the ion.
These tasks are solved by adapting a 'stylus trap'~\cite{maiwald2009}
and combining it with a parabolic mirror.
The aluminium parabolic mirror replaces the planar ground electrode of
the stylus trap~\cite{maiwald2009,maiwald2012}. 
Two tubular electrodes providing ground potential and the radio-frequency
signal, respectively, are mounted concentrically and attached to a
piezo-driven translation stage.
The latter allows for adjusting the minimum of the trapping potential
with nm-accuracy relative to the mirror's focus.
The layout is sketched in Fig.~\ref{fig.trap}.
The level scheme relevant for laser cooling YbII and for the coupling
experiments outlined in Sec.\,\ref{sec.efficiency} is depicted in
Fig.~\ref{fig.Yb}. 

The radiation from a single YbII ion's fluorescence
as collimated by the parabolic mirror is
displayed in Fig.~\ref{fig.trap}b. 
The cooling transition has been saturated in this experiment.
The figure shows an image of the intensity distribution at the
output plane of the paraboloid.
The spatial distribution of the measured fluorescence photons
corresponds to the one of an isotropic point source, as which a
saturated YbII ion can be considered~\cite{maiwald2012}.
The thin, concentric, ring-shaped features originate from surface
distortions of the parabolic mirror, see discussion below and in
Sec.~\ref{sec.efficiency}. 
The central dark spot is attributed to the opening for the trap electrodes.

\subsection{Fluorescence collection}

The large solid angle covered by a deep parabolic mirror is not only
beneficial for focusing onto an atom, but also -- as suggested by the
time-reversal symmetry arguments raised before -- for the efficient
collection of photons emerging from the atom. 
The latter is of importance in any application in which the internal
state of an atom has to be determined.

For our mirror geometry we  demonstrated the collection of 54\% of
the emitted fluorescence photons~\cite{maiwald2012}.
This number is basically limited by two constraints:
As mentioned above, a saturated YbII ion emits on average as an
isotropic point source.
For the latter the solid angle covered by our mirror amounts to 81\%
instead of the 94\% when considering the emission of a linear dipole
aligned with the mirror's optical axis.
The second constraint stems from the surface quality of the used
mirror.
The nominal reflectivity of aluminium, when averaging over all
angles of incidence corresponding to the surface of our mirror, should
amount to roughly 87\%.
Instead, the measured reflectivity amounts to 67\%.
Such a low value is mainly attributed to a too large surface roughness
of the parabolic mirror, which might be caused by a non-optimum
manufacturing process.
Using a mirror with better surface properties along with collecting
from a linear-dipole emitter should boost the collection efficiency to
82\%.

Nevertheless, according to current literature the number of
two million photons per second actually counted in our setup outperforms
other setups using neutral atoms or ions.

\section{Experimental determination of the coupling efficiency}
\label{sec.efficiency}

With the different aspects treated so far we now have basically all
tools at hand for conducting experiments on efficient light-matter
coupling.
What we did not discuss, however, is a reliable way of determining the
coupling efficiency $G$ experimentally.
In principle, one could conduct the experiment one would like to do
(e.g. phase shift or extinction measurements) and compare the achieved
results with theory in order to find a reasonable value for $G$ that
consistently reproduces the experiment.
But there are some obstacles.
As both the phase shift and the extinction induced by a single atom
are caused by the interference of incident and scattered light, any
incoherent scattering due to saturation of the atom reduces the
corresponding figure of merit~\cite{wrigge2008,pototschnig2011,
  sondermann2013p}.
Also the amount of light scattered back into the solid-angle
cone of the focusing optics, as measured in Ref.~\cite{aljunid2011},
is influenced by the saturation of the atom's excited state.
This is due to the fact that the scattering ratio is decreasing for
increasing saturation parameter~\cite{sondermann2013p,vanenk2004}.

One could solve such problems by determining the atomic saturation in
an additional experiment.
But there is a simpler solution in measuring \emph{only} a
saturation curve and relating the power needed to achieve a certain
saturation parameter $S$ to the one necessary under ideal conditions,
i.e. at $G=1$~\cite{fischer2014}.
The saturation parameter $S$ induced by light with detuning $\Delta$
and incident power $P$ is given by~\cite{sondermann2013p}
\begin{equation}
\label{eq.saturation}
S= G\cdot\frac{8P}{\hbar\omega_0\Gamma}\cdot
\frac{1}{1+4\Delta^2/\Gamma^2}\quad ,
\end{equation}
where $\omega_0$ is the resonance frequency of the atomic transition.
For example, the minimum power ($G=1$) to achieve a unit saturation
parameter on resonance is $\hbar\omega_0\Gamma/8$, see also
Ref.~\cite{kochan1994}.

In the experiment one measures the amount of fluorescence counts as a
function of the incident power and fits the result to a function
proportional to 
\begin{equation}
\label{eq.satcurve}
\frac{\Gamma}{2}\cdot \frac{S(P)}{1+S(P)}\quad, 
\end{equation}
which is the rate of photons
scattered by a two-level atom in the steady state.
The only difficulty arises from distinguishing the scattered photons
from the incident ones.
This could be done by monitoring e.g. the Stokes-shifted fluorescence
when coupling to a molecule~\cite{wrigge2008}.
Working with YbII ions one could monitor the photons emitted on the
auxiliary transition $^3[3/2]_{1/2}\rightarrow\ ^2S_{1/2}$ at 297\,nm
which is part of the typical scheme applied for laser-cooling
YbII~\cite{bell1991}.

\begin{figure}
\includegraphics{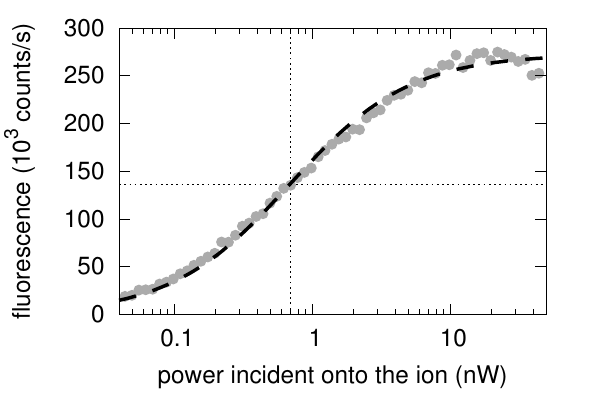}
\caption{\label{fig.saturation}
Saturation curve (symbols) obtained cooling an YbII ion with a
doughnut mode incident from $r\le2f$ at a detuning
$\Delta=\Gamma/2$.
The dashed line denotes the result of fitting
Eq.~\ref{eq.satcurve} to the experimental data.
The dotted lines indicate the power and fluorescence rate at
$S=1$.}
\end{figure}

Here we pursue another method and split incident and scattered light
spatially.
The spatial separation is accomplished by restricting the incident
light to half of the solid angle as depicted in the extinction setup
in Fig.~\ref{fig.extinction}.
In other words, we cool the YbII ion with a radially polarized
doughnut mode focused by the parabolic mirror while restricting the
incident light to radii 
$r\le2f$.\footnote{The cooling laser beam at 370~\,nm that enters
  through an auxiliary opening of the parabolic mirror
  (cf. Fig.~\ref{fig.trap}) is blocked during the saturation measurements.}
Furthermore, one has to use the same kind of aperture in the detection
path as in the excitation path in order to detect only backward scattered
light~\cite{fischer2014}.
The result of such an experiment is given in Fig.~\ref{fig.saturation}.

A least-squares fit yields a power of 690\,pW for achieving $S=1$.
Using the parameters $\Gamma^{-1}=$ 8.1\,ns, $\Delta=\Gamma/2$ and
$\omega_0=2\pi c_0/$370\,nm Eq.~\ref{eq.saturation} thus delivers
$G=0.024$. 
However, we have to take into account that Eq.~\ref{eq.saturation} is
valid for a two-level atom, which is not the appropriate description
of YbII. 
The focused radially polarized mode only drives the $\pi$-transition
which has a relative oscillator strength of $1/3$ in comparison to
$2/3$ for the $\sigma_\pm$-transitions for the levels $S_{1/2}$ and
$P_{1/2}$.
Hence we have to apply a correction factor of three in order to obtain
$G=0.072$.\footnote{This reasoning assumes that the quantization axis is
  parallel to the optical axis of the parabola. But one can show that
  for any orientation of the quantization axis the same correction factor
  has to be applied when treating a $S_{1/2}\rightarrow P_{1/2}$ transition.}

This coupling efficiency is among the largest achieved in a
free-space setup so far, but it is seven times below the expected value: 
Focusing from half solid angle, the maximum achievable coupling efficiency
amounts to $G=0.5$.
The non-perfect overlap of $\eta=0.98$ measured for the incident
doughnut modes is so close to the ideal value that it can only explain
a tiny portion of the discrepancy.
Most probably the discrepancy arises from the fact that the
aberrations of the non-perfect parabolic surface have not been
compensated during the measurement.
From an interferometric characterization\cite{leuchs2008} of the
parabolic mirror performed before mounting it in the vacuum chamber
and simulations of the focal intensity based on the corresponding
results we predict a Strehl ratio of 87\% for focusing from $r\le2f$.
The obtainable coupling efficiency including the aberrations and
$\eta$ thus amounts to 41\%. 

This value is still much larger than the measured one.
We speculate that the surface of the parabolic mirror might be
subjected to unknown distortions which are not recognizable in the
interferometric measurements performed at visible wavelengths.
But also distortions of the incident wavefront by the viewport of the
vacuum chamber might play a role.
In any case, the measurement of the coupling efficiency as performed
here provides a sensitive tool hinting at any open issues.

\section{Outlook}

The experimental demonstrations of coupling light to a single atom in
free space have been made with different systems.
The best experimental performance in relation to the phase shift of
the full transmitted beam, the extinction of the irradiating beam and the
absorption of a single photon are are still far from the theoretical best
possible values and thus far away from the maximum possible
efficiency.
The parabolic mirror set-up emphasized here offers the opportunity of
improving upon all these numbers. 
Nevertheless, demonstrating close to 100\% absorption efficiency
remains a particular challenge, since the requirements on
mode-matching in the spatial and temporal domain have to be fulfilled
with highest quality. 
This is especially difficult at the short wavelength of the
linear-dipole transition of the almost ideal two-level system YbIII,
which we successfully trapped recently.
We hope to report on light-matter-interaction experiments with a
single YbIII ion in the near future.

\begin{acknowledgments}
We gratefully acknowledge the contributions of Marianne Bader, 
Benoit Chalopin, Martin Fischer, Andrea Golla, Simon Heugel 
and Robert Maiwald to our experimental endeavours.
We thank the \emph{Deutsche Forschungsgemeinschaft} for financial
support. 
G.L. also acknowledges financial support from the \emph{European Research
  Council} under the Advanced Grant \lq PACART\rq.
\end{acknowledgments}

%%%%%%%%%%%%%%%%%%%%%%%%%%%%%%%%%%%%%%%%%%%%%%%%%%%%%%%%%%%%%
%% \bibliographystyle{apsrev4-1}
%% \bibliography{/home/markus/Dokumente/bibtex/optik}

\begin{thebibliography}{78}%
\makeatletter
\providecommand \@ifxundefined [1]{%
 \@ifx{#1\undefined}
}%
\providecommand \@ifnum [1]{%
 \ifnum #1\expandafter \@firstoftwo
 \else \expandafter \@secondoftwo
 \fi
}%
\providecommand \@ifx [1]{%
 \ifx #1\expandafter \@firstoftwo
 \else \expandafter \@secondoftwo
 \fi
}%
\providecommand \natexlab [1]{#1}%
\providecommand \enquote  [1]{``#1''}%
\providecommand \bibnamefont  [1]{#1}%
\providecommand \bibfnamefont [1]{#1}%
\providecommand \citenamefont [1]{#1}%
\providecommand \href@noop [0]{\@secondoftwo}%
\providecommand \href [0]{\begingroup \@sanitize@url \@href}%
\providecommand \@href[1]{\@@startlink{#1}\@@href}%
\providecommand \@@href[1]{\endgroup#1\@@endlink}%
\providecommand \@sanitize@url [0]{\catcode `\\12\catcode `\$12\catcode
  `\&12\catcode `\#12\catcode `\^12\catcode `\_12\catcode `\%12\relax}%
\providecommand \@@startlink[1]{}%
\providecommand \@@endlink[0]{}%
\providecommand \url  [0]{\begingroup\@sanitize@url \@url }%
\providecommand \@url [1]{\endgroup\@href {#1}{\urlprefix }}%
\providecommand \urlprefix  [0]{URL }%
\providecommand \Eprint [0]{\href }%
\providecommand \doibase [0]{http://dx.doi.org/}%
\providecommand \selectlanguage [0]{\@gobble}%
\providecommand \bibinfo  [0]{\@secondoftwo}%
\providecommand \bibfield  [0]{\@secondoftwo}%
\providecommand \translation [1]{[#1]}%
\providecommand \BibitemOpen [0]{}%
\providecommand \bibitemStop [0]{}%
\providecommand \bibitemNoStop [0]{.\EOS\space}%
\providecommand \EOS [0]{\spacefactor3000\relax}%
\providecommand \BibitemShut  [1]{\csname bibitem#1\endcsname}%
\let\auto@bib@innerbib\@empty
%</preamble>
\bibitem [{\citenamefont {Mandel}\ and\ \citenamefont
  {Wolf}(1995)}]{mandel-wolf1995}%
  \BibitemOpen
  \bibfield  {author} {\bibinfo {author} {\bibfnamefont {L.}~\bibnamefont
  {Mandel}}\ and\ \bibinfo {author} {\bibfnamefont {E.}~\bibnamefont {Wolf}},\
  }\href@noop {} {\emph {\bibinfo {title} {Optical Coherence and Quantum
  Optics}}}\ (\bibinfo  {publisher} {Cambridge University Press},\ \bibinfo
  {address} {Cambridge, New York},\ \bibinfo {year} {1995})\BibitemShut
  {NoStop}%
\bibitem [{\citenamefont {Scully}\ and\ \citenamefont
  {Zubairy}(1997)}]{scully1997}%
  \BibitemOpen
  \bibfield  {author} {\bibinfo {author} {\bibfnamefont {M.~O.}\ \bibnamefont
  {Scully}}\ and\ \bibinfo {author} {\bibfnamefont {M.~S.}\ \bibnamefont
  {Zubairy}},\ }\href@noop {} {\emph {\bibinfo {title} {Quantum {O}ptics}}}\
  (\bibinfo  {publisher} {Cambridge University Press},\ \bibinfo {address}
  {Cambridge},\ \bibinfo {year} {1997})\BibitemShut {NoStop}%
\bibitem [{\citenamefont {Moskovits}(1985)}]{moskovits1985}%
  \BibitemOpen
  \bibfield  {author} {\bibinfo {author} {\bibfnamefont {M.}~\bibnamefont
  {Moskovits}},\ }\href {\doibase 10.1103/RevModPhys.57.783} {\bibfield
  {journal} {\bibinfo  {journal} {Rev. Mod. Phys.}\ }\textbf {\bibinfo {volume}
  {57}},\ \bibinfo {pages} {783} (\bibinfo {year} {1985})}\BibitemShut
  {NoStop}%
\bibitem [{\citenamefont {K\"uhn}\ \emph {et~al.}(2006)\citenamefont {K\"uhn},
  \citenamefont {H\aa{}kanson}, \citenamefont {Rogobete},\ and\ \citenamefont
  {Sandoghdar}}]{kuehn2006}%
  \BibitemOpen
  \bibfield  {author} {\bibinfo {author} {\bibfnamefont {S.}~\bibnamefont
  {K\"uhn}}, \bibinfo {author} {\bibfnamefont {U.}~\bibnamefont
  {H\aa{}kanson}}, \bibinfo {author} {\bibfnamefont {L.}~\bibnamefont
  {Rogobete}}, \ and\ \bibinfo {author} {\bibfnamefont {V.}~\bibnamefont
  {Sandoghdar}},\ }\href {\doibase 10.1103/PhysRevLett.97.017402} {\bibfield
  {journal} {\bibinfo  {journal} {Phys. Rev. Lett.}\ }\textbf {\bibinfo
  {volume} {97}},\ \bibinfo {pages} {017402} (\bibinfo {year}
  {2006})}\BibitemShut {NoStop}%
\bibitem [{\citenamefont {Novotny}\ and\ \citenamefont {van
  Hulst}(2011)}]{novotny2011}%
  \BibitemOpen
  \bibfield  {author} {\bibinfo {author} {\bibfnamefont {L.}~\bibnamefont
  {Novotny}}\ and\ \bibinfo {author} {\bibfnamefont {N.}~\bibnamefont {van
  Hulst}},\ }\href@noop {} {\bibfield  {journal} {\bibinfo  {journal} {Nature
  Photonics}\ }\textbf {\bibinfo {volume} {5}},\ \bibinfo {pages} {83}
  (\bibinfo {year} {2011})}\BibitemShut {NoStop}%
\bibitem [{\citenamefont {Raimond}\ \emph {et~al.}(2001)\citenamefont
  {Raimond}, \citenamefont {Brune},\ and\ \citenamefont
  {Haroche}}]{raimond2001}%
  \BibitemOpen
  \bibfield  {author} {\bibinfo {author} {\bibfnamefont {J.~M.}\ \bibnamefont
  {Raimond}}, \bibinfo {author} {\bibfnamefont {M.}~\bibnamefont {Brune}}, \
  and\ \bibinfo {author} {\bibfnamefont {S.}~\bibnamefont {Haroche}},\ }\href
  {\doibase 10.1103/RevModPhys.73.565} {\bibfield  {journal} {\bibinfo
  {journal} {Rev. Mod. Phys.}\ }\textbf {\bibinfo {volume} {73}},\ \bibinfo
  {pages} {565} (\bibinfo {year} {2001})}\BibitemShut {NoStop}%
\bibitem [{\citenamefont {Walther}\ \emph {et~al.}(2006)\citenamefont
  {Walther}, \citenamefont {Varcoe}, \citenamefont {Englert},\ and\
  \citenamefont {Becker}}]{walther2006}%
  \BibitemOpen
  \bibfield  {author} {\bibinfo {author} {\bibfnamefont {H.}~\bibnamefont
  {Walther}}, \bibinfo {author} {\bibfnamefont {B.~T.~H.}\ \bibnamefont
  {Varcoe}}, \bibinfo {author} {\bibfnamefont {B.-G.}\ \bibnamefont {Englert}},
  \ and\ \bibinfo {author} {\bibfnamefont {T.}~\bibnamefont {Becker}},\
  }\href@noop {} {\bibfield  {journal} {\bibinfo  {journal} {Rep. Prog. Phys.}\
  }\textbf {\bibinfo {volume} {69}},\ \bibinfo {pages} {1325} (\bibinfo {year}
  {2006})}\BibitemShut {NoStop}%
\bibitem [{\citenamefont {Kimble}(1998)}]{kimble1998}%
  \BibitemOpen
  \bibfield  {author} {\bibinfo {author} {\bibfnamefont {H.~J.}\ \bibnamefont
  {Kimble}},\ }\href@noop {} {\bibfield  {journal} {\bibinfo  {journal}
  {Physica Scripta}\ }\textbf {\bibinfo {volume} {T76}},\ \bibinfo {pages}
  {127} (\bibinfo {year} {1998})}\BibitemShut {NoStop}%
\bibitem [{\citenamefont {Rempe}(1993)}]{rempe1993}%
  \BibitemOpen
  \bibfield  {author} {\bibinfo {author} {\bibfnamefont {G.}~\bibnamefont
  {Rempe}},\ }\href@noop {} {\bibfield  {journal} {\bibinfo  {journal}
  {Contemporary Physics}\ }\textbf {\bibinfo {volume} {34}},\ \bibinfo {pages}
  {119} (\bibinfo {year} {1993})}\BibitemShut {NoStop}%
\bibitem [{\citenamefont {Sondermann}\ \emph {et~al.}(2007)\citenamefont
  {Sondermann}, \citenamefont {Maiwald}, \citenamefont {Konermann},
  \citenamefont {Lindlein}, \citenamefont {Peschel},\ and\ \citenamefont
  {Leuchs}}]{sondermann2007}%
  \BibitemOpen
  \bibfield  {author} {\bibinfo {author} {\bibfnamefont {M.}~\bibnamefont
  {Sondermann}}, \bibinfo {author} {\bibfnamefont {R.}~\bibnamefont {Maiwald}},
  \bibinfo {author} {\bibfnamefont {H.}~\bibnamefont {Konermann}}, \bibinfo
  {author} {\bibfnamefont {N.}~\bibnamefont {Lindlein}}, \bibinfo {author}
  {\bibfnamefont {U.}~\bibnamefont {Peschel}}, \ and\ \bibinfo {author}
  {\bibfnamefont {G.}~\bibnamefont {Leuchs}},\ }\href {\doibase
  10.1007/s00340-007-2859-4} {\bibfield  {journal} {\bibinfo  {journal} {Appl.
  Phys. B}\ }\textbf {\bibinfo {volume} {89}},\ \bibinfo {pages} {489}
  (\bibinfo {year} {2007})},\ \Eprint {http://arxiv.org/abs/arxiv:0708.0772}
  {arxiv:0708.0772} \BibitemShut {NoStop}%
\bibitem [{\citenamefont {Specht}\ \emph {et~al.}(2011)\citenamefont {Specht},
  \citenamefont {N{\"o}lleke}, \citenamefont {Reiserer}, \citenamefont
  {Uphoff}, \citenamefont {Figueroa}, \citenamefont {Ritter},\ and\
  \citenamefont {Rempe}}]{specht2011}%
  \BibitemOpen
  \bibfield  {author} {\bibinfo {author} {\bibfnamefont {H.}~\bibnamefont
  {Specht}}, \bibinfo {author} {\bibfnamefont {C.}~\bibnamefont {N{\"o}lleke}},
  \bibinfo {author} {\bibfnamefont {A.}~\bibnamefont {Reiserer}}, \bibinfo
  {author} {\bibfnamefont {M.}~\bibnamefont {Uphoff}}, \bibinfo {author}
  {\bibfnamefont {E.}~\bibnamefont {Figueroa}}, \bibinfo {author}
  {\bibfnamefont {S.}~\bibnamefont {Ritter}}, \ and\ \bibinfo {author}
  {\bibfnamefont {G.}~\bibnamefont {Rempe}},\ }\href@noop {} {\bibfield
  {journal} {\bibinfo  {journal} {Nature}\ }\textbf {\bibinfo {volume} {473}},\
  \bibinfo {pages} {190} (\bibinfo {year} {2011})}\BibitemShut {NoStop}%
\bibitem [{\citenamefont {Ritter}\ \emph {et~al.}(2012)\citenamefont {Ritter},
  \citenamefont {N\"{o}lleke}, \citenamefont {Hahn}, \citenamefont {Reiserer},
  \citenamefont {Neuzner}, \citenamefont {Uphoff}, \citenamefont {M\"{u}cke},
  \citenamefont {Figueroa}, \citenamefont {Bochmann},\ and\ \citenamefont
  {Rempe}}]{ritter2012}%
  \BibitemOpen
  \bibfield  {author} {\bibinfo {author} {\bibfnamefont {S.}~\bibnamefont
  {Ritter}}, \bibinfo {author} {\bibfnamefont {C.}~\bibnamefont {N\"{o}lleke}},
  \bibinfo {author} {\bibfnamefont {C.}~\bibnamefont {Hahn}}, \bibinfo {author}
  {\bibfnamefont {A.}~\bibnamefont {Reiserer}}, \bibinfo {author}
  {\bibfnamefont {A.}~\bibnamefont {Neuzner}}, \bibinfo {author} {\bibfnamefont
  {M.}~\bibnamefont {Uphoff}}, \bibinfo {author} {\bibfnamefont
  {M.}~\bibnamefont {M\"{u}cke}}, \bibinfo {author} {\bibfnamefont
  {E.}~\bibnamefont {Figueroa}}, \bibinfo {author} {\bibfnamefont
  {J.}~\bibnamefont {Bochmann}}, \ and\ \bibinfo {author} {\bibfnamefont
  {G.}~\bibnamefont {Rempe}},\ }\href {\doibase 10.1038/nature11023} {\bibfield
   {journal} {\bibinfo  {journal} {Nature}\ }\textbf {\bibinfo {volume}
  {484}},\ \bibinfo {pages} {195} (\bibinfo {year} {2012})},\ \Eprint
  {http://arxiv.org/abs/arXiv:1202.5955} {arXiv:1202.5955} \BibitemShut
  {NoStop}%
\bibitem [{\citenamefont {Sames}\ \emph {et~al.}(2014)\citenamefont {Sames},
  \citenamefont {Chibani}, \citenamefont {Hamsen}, \citenamefont {Altin},
  \citenamefont {Wilk},\ and\ \citenamefont {Rempe}}]{sames2014}%
  \BibitemOpen
  \bibfield  {author} {\bibinfo {author} {\bibfnamefont {C.}~\bibnamefont
  {Sames}}, \bibinfo {author} {\bibfnamefont {H.}~\bibnamefont {Chibani}},
  \bibinfo {author} {\bibfnamefont {C.}~\bibnamefont {Hamsen}}, \bibinfo
  {author} {\bibfnamefont {P.~A.}\ \bibnamefont {Altin}}, \bibinfo {author}
  {\bibfnamefont {T.}~\bibnamefont {Wilk}}, \ and\ \bibinfo {author}
  {\bibfnamefont {G.}~\bibnamefont {Rempe}},\ }\href {\doibase
  10.1103/PhysRevLett.112.043601} {\bibfield  {journal} {\bibinfo  {journal}
  {Phys. Rev. Lett.}\ }\textbf {\bibinfo {volume} {112}},\ \bibinfo {pages}
  {043601} (\bibinfo {year} {2014})}\BibitemShut {NoStop}%
\bibitem [{\citenamefont {Tiecke}\ \emph {et~al.}(2014)\citenamefont {Tiecke},
  \citenamefont {Thompson}, \citenamefont {de~Leon}, \citenamefont {Liu},
  \citenamefont {Vuleti{\'c}},\ and\ \citenamefont {Lukin}}]{tiecke2014}%
  \BibitemOpen
  \bibfield  {author} {\bibinfo {author} {\bibfnamefont {T.}~\bibnamefont
  {Tiecke}}, \bibinfo {author} {\bibfnamefont {J.}~\bibnamefont {Thompson}},
  \bibinfo {author} {\bibfnamefont {N.}~\bibnamefont {de~Leon}}, \bibinfo
  {author} {\bibfnamefont {L.}~\bibnamefont {Liu}}, \bibinfo {author}
  {\bibfnamefont {V.}~\bibnamefont {Vuleti{\'c}}}, \ and\ \bibinfo {author}
  {\bibfnamefont {M.}~\bibnamefont {Lukin}},\ }\href {\doibase
  10.1038/nature13188} {\bibfield  {journal} {\bibinfo  {journal} {Nature}\
  }\textbf {\bibinfo {volume} {508}},\ \bibinfo {pages} {241} (\bibinfo {year}
  {2014})}\BibitemShut {NoStop}%
\bibitem [{\citenamefont {Quabis}\ \emph {et~al.}(2000)\citenamefont {Quabis},
  \citenamefont {Dorn}, \citenamefont {Eberler}, \citenamefont {Gl\"ockl},\
  and\ \citenamefont {Leuchs}}]{quabis2000}%
  \BibitemOpen
  \bibfield  {author} {\bibinfo {author} {\bibfnamefont {S.}~\bibnamefont
  {Quabis}}, \bibinfo {author} {\bibfnamefont {R.}~\bibnamefont {Dorn}},
  \bibinfo {author} {\bibfnamefont {M.}~\bibnamefont {Eberler}}, \bibinfo
  {author} {\bibfnamefont {O.}~\bibnamefont {Gl\"ockl}}, \ and\ \bibinfo
  {author} {\bibfnamefont {G.}~\bibnamefont {Leuchs}},\ }\href@noop {}
  {\bibfield  {journal} {\bibinfo  {journal} {Opt. Comm.}\ }\textbf {\bibinfo
  {volume} {179}},\ \bibinfo {pages} {1} (\bibinfo {year} {2000})}\BibitemShut
  {NoStop}%
\bibitem [{\citenamefont {Leuchs}\ and\ \citenamefont
  {Sondermann}(2012)}]{leuchs2012}%
  \BibitemOpen
  \bibfield  {author} {\bibinfo {author} {\bibfnamefont {G.}~\bibnamefont
  {Leuchs}}\ and\ \bibinfo {author} {\bibfnamefont {M.}~\bibnamefont
  {Sondermann}},\ }\href {\doibase 10.1088/0031-8949/85/05/058101} {\bibfield
  {journal} {\bibinfo  {journal} {Physica Scripta}\ }\textbf {\bibinfo {volume}
  {85}},\ \bibinfo {pages} {058101} (\bibinfo {year} {2012})},\ \Eprint
  {http://arxiv.org/abs/arXiv:1205.1374} {arXiv:1205.1374} \BibitemShut
  {NoStop}%
\bibitem [{\citenamefont {Leuchs}\ and\ \citenamefont
  {Sondermann}(2013)}]{leuchs2013o}%
  \BibitemOpen
  \bibfield  {author} {\bibinfo {author} {\bibfnamefont {G.}~\bibnamefont
  {Leuchs}}\ and\ \bibinfo {author} {\bibfnamefont {M.}~\bibnamefont
  {Sondermann}},\ }\href {\doibase 10.1080/09500340.2012.716461} {\bibfield
  {journal} {\bibinfo  {journal} {Journal of Modern Optics}\ }\textbf {\bibinfo
  {volume} {60}},\ \bibinfo {pages} {36} (\bibinfo {year} {2013})}\BibitemShut
  {NoStop}%
\bibitem [{\citenamefont {Silberfarb}\ and\ \citenamefont
  {Deutsch}(2003)}]{silberfarb2003}%
  \BibitemOpen
  \bibfield  {author} {\bibinfo {author} {\bibfnamefont {A.}~\bibnamefont
  {Silberfarb}}\ and\ \bibinfo {author} {\bibfnamefont {I.~H.}\ \bibnamefont
  {Deutsch}},\ }\href {\doibase 10.1103/PhysRevA.68.013817} {\bibfield
  {journal} {\bibinfo  {journal} {Phys. Rev. A}\ }\textbf {\bibinfo {volume}
  {68}},\ \bibinfo {pages} {013817} (\bibinfo {year} {2003})}\BibitemShut
  {NoStop}%
\bibitem [{\citenamefont {Pinotsi}\ and\ \citenamefont
  {Imamoglu}(2008)}]{pinotsi2008}%
  \BibitemOpen
  \bibfield  {author} {\bibinfo {author} {\bibfnamefont {D.}~\bibnamefont
  {Pinotsi}}\ and\ \bibinfo {author} {\bibfnamefont {A.}~\bibnamefont
  {Imamoglu}},\ }\href {\doibase 10.1103/PhysRevLett.100.093603} {\bibfield
  {journal} {\bibinfo  {journal} {Phys. Rev. Lett.}\ }\textbf {\bibinfo
  {volume} {100}},\ \bibinfo {pages} {093603} (\bibinfo {year}
  {2008})}\BibitemShut {NoStop}%
\bibitem [{\citenamefont {Cohen-Tannoudji}\ \emph {et~al.}(1989)\citenamefont
  {Cohen-Tannoudji}, \citenamefont {Dupont-Roc},\ and\ \citenamefont
  {Grynberg}}]{cohen-tannoudji1989}%
  \BibitemOpen
  \bibfield  {author} {\bibinfo {author} {\bibfnamefont {C.}~\bibnamefont
  {Cohen-Tannoudji}}, \bibinfo {author} {\bibfnamefont {J.}~\bibnamefont
  {Dupont-Roc}}, \ and\ \bibinfo {author} {\bibfnamefont {G.}~\bibnamefont
  {Grynberg}},\ }\href@noop {} {\emph {\bibinfo {title} {Photons and atoms}}}\
  (\bibinfo  {publisher} {J. Wiley and Sons},\ \bibinfo {year}
  {1989})\BibitemShut {NoStop}%
\bibitem [{\citenamefont {Basset}(1986)}]{basset1986}%
  \BibitemOpen
  \bibfield  {author} {\bibinfo {author} {\bibfnamefont {I.~M.}\ \bibnamefont
  {Basset}},\ }\href@noop {} {\bibfield  {journal} {\bibinfo  {journal}
  {Journal of Modern Optics}\ }\textbf {\bibinfo {volume} {33}},\ \bibinfo
  {pages} {279} (\bibinfo {year} {1986})}\BibitemShut {NoStop}%
\bibitem [{\citenamefont {van Enk}(2004)}]{vanenk2004}%
  \BibitemOpen
  \bibfield  {author} {\bibinfo {author} {\bibfnamefont {S.~J.}\ \bibnamefont
  {van Enk}},\ }\href@noop {} {\bibfield  {journal} {\bibinfo  {journal} {Phys.
  Rev. A}\ }\textbf {\bibinfo {volume} {69}},\ \bibinfo {pages} {043813}
  (\bibinfo {year} {2004})}\BibitemShut {NoStop}%
\bibitem [{\citenamefont {Wang}\ \emph {et~al.}(2011)\citenamefont {Wang},
  \citenamefont {Min\'a\v{r}}, \citenamefont {Sheridan},\ and\ \citenamefont
  {Scarani}}]{wang2011}%
  \BibitemOpen
  \bibfield  {author} {\bibinfo {author} {\bibfnamefont {Y.}~\bibnamefont
  {Wang}}, \bibinfo {author} {\bibfnamefont {J.}~\bibnamefont {Min\'a\v{r}}},
  \bibinfo {author} {\bibfnamefont {L.}~\bibnamefont {Sheridan}}, \ and\
  \bibinfo {author} {\bibfnamefont {V.}~\bibnamefont {Scarani}},\ }\href
  {\doibase 10.1103/PhysRevA.83.063842} {\bibfield  {journal} {\bibinfo
  {journal} {Phys. Rev. A}\ }\textbf {\bibinfo {volume} {83}},\ \bibinfo
  {pages} {063842} (\bibinfo {year} {2011})}\BibitemShut {NoStop}%
\bibitem [{\citenamefont {Sondermann}\ \emph {et~al.}(2008)\citenamefont
  {Sondermann}, \citenamefont {Lindlein},\ and\ \citenamefont
  {Leuchs}}]{sondermann2008}%
  \BibitemOpen
  \bibfield  {author} {\bibinfo {author} {\bibfnamefont {M.}~\bibnamefont
  {Sondermann}}, \bibinfo {author} {\bibfnamefont {N.}~\bibnamefont
  {Lindlein}}, \ and\ \bibinfo {author} {\bibfnamefont {G.}~\bibnamefont
  {Leuchs}},\ }\href@noop {} {\bibfield  {journal} {\bibinfo  {journal}
  \Eprint {http://arxiv.org/abs/arXiv:0811.2098} {arXiv:0811.2098
    [physics.optics]}\ } (\bibinfo {year} {2008})}\BibitemShut 
  {NoStop}%
\bibitem [{\citenamefont {Jackson}(1999)}]{jackson1999}%
  \BibitemOpen
  \bibfield  {author} {\bibinfo {author} {\bibfnamefont {J.~D.}\ \bibnamefont
  {Jackson}},\ }\href@noop {} {\emph {\bibinfo {title} {Classical
  Electrodynamics}}},\ \bibinfo {edition} {3rd}\ ed.\ (\bibinfo  {publisher}
  {Wiley, New York},\ \bibinfo {year} {1999})\BibitemShut {NoStop}%
\bibitem [{\citenamefont {Fischer}\ \emph {et~al.}(2014)\citenamefont
  {Fischer}, \citenamefont {Bader}, \citenamefont {Maiwald}, \citenamefont
  {Golla}, \citenamefont {Sondermann},\ and\ \citenamefont
  {Leuchs}}]{fischer2014}%
  \BibitemOpen
  \bibfield  {author} {\bibinfo {author} {\bibfnamefont {M.}~\bibnamefont
  {Fischer}}, \bibinfo {author} {\bibfnamefont {M.}~\bibnamefont {Bader}},
  \bibinfo {author} {\bibfnamefont {R.}~\bibnamefont {Maiwald}}, \bibinfo
  {author} {\bibfnamefont {A.}~\bibnamefont {Golla}}, \bibinfo {author}
  {\bibfnamefont {M.}~\bibnamefont {Sondermann}}, \ and\ \bibinfo {author}
  {\bibfnamefont {G.}~\bibnamefont {Leuchs}},\ }\href {\doibase
  10.1007/s00340-014-5817-y} {\bibfield  {journal} {\bibinfo  {journal} {Appl.
  Phys. B}\ }\textbf {\bibinfo {volume} {117}},\ \bibinfo {pages} {797}
  (\bibinfo {year} {2014})},\ \Eprint {http://arxiv.org/abs/arXiv:1311.1982}
  {arXiv:1311.1982} \BibitemShut {NoStop}%
\bibitem [{\citenamefont {Lindlein}\ \emph {et~al.}(2007)\citenamefont
  {Lindlein}, \citenamefont {Maiwald}, \citenamefont {Konermann}, \citenamefont
  {Sondermann}, \citenamefont {Peschel},\ and\ \citenamefont
  {Leuchs}}]{lindlein2007}%
  \BibitemOpen
  \bibfield  {author} {\bibinfo {author} {\bibfnamefont {N.}~\bibnamefont
  {Lindlein}}, \bibinfo {author} {\bibfnamefont {R.}~\bibnamefont {Maiwald}},
  \bibinfo {author} {\bibfnamefont {H.}~\bibnamefont {Konermann}}, \bibinfo
  {author} {\bibfnamefont {M.}~\bibnamefont {Sondermann}}, \bibinfo {author}
  {\bibfnamefont {U.}~\bibnamefont {Peschel}}, \ and\ \bibinfo {author}
  {\bibfnamefont {G.}~\bibnamefont {Leuchs}},\ }\href {\doibase
  10.1134/S1054660X07070055} {\bibfield  {journal} {\bibinfo  {journal} {Laser
  Physics}\ }\textbf {\bibinfo {volume} {17}},\ \bibinfo {pages} {927}
  (\bibinfo {year} {2007})}\BibitemShut {NoStop}%
\bibitem [{\citenamefont {Golla}\ \emph {et~al.}(2012)\citenamefont {Golla},
  \citenamefont {Chalopin}, \citenamefont {Bader}, \citenamefont {Harder},
  \citenamefont {Mantel}, \citenamefont {Maiwald}, \citenamefont {Lindlein},
  \citenamefont {Sondermann},\ and\ \citenamefont {Leuchs}}]{golla2012}%
  \BibitemOpen
  \bibfield  {author} {\bibinfo {author} {\bibfnamefont {A.}~\bibnamefont
  {Golla}}, \bibinfo {author} {\bibfnamefont {B.}~\bibnamefont {Chalopin}},
  \bibinfo {author} {\bibfnamefont {M.}~\bibnamefont {Bader}}, \bibinfo
  {author} {\bibfnamefont {I.}~\bibnamefont {Harder}}, \bibinfo {author}
  {\bibfnamefont {K.}~\bibnamefont {Mantel}}, \bibinfo {author} {\bibfnamefont
  {R.}~\bibnamefont {Maiwald}}, \bibinfo {author} {\bibfnamefont
  {N.}~\bibnamefont {Lindlein}}, \bibinfo {author} {\bibfnamefont
  {M.}~\bibnamefont {Sondermann}}, \ and\ \bibinfo {author} {\bibfnamefont
  {G.}~\bibnamefont {Leuchs}},\ }\href {\doibase 10.1140/epjd/e2012-30293-y}
  {\bibfield  {journal} {\bibinfo  {journal} {Eur. Phys. J. D}\ }\textbf
  {\bibinfo {volume} {66}},\ \bibinfo {pages} {190} (\bibinfo {year} {2012})},\
  \Eprint {http://arxiv.org/abs/arXiv:1207.3215} {arXiv:1207.3215} \BibitemShut
  {NoStop}%
\bibitem [{\citenamefont {Hell}\ and\ \citenamefont
  {Stelzer}(1992)}]{hell1992}%
  \BibitemOpen
  \bibfield  {author} {\bibinfo {author} {\bibfnamefont {S.}~\bibnamefont
  {Hell}}\ and\ \bibinfo {author} {\bibfnamefont {E.~H.~K.}\ \bibnamefont
  {Stelzer}},\ }\href@noop {} {\bibfield  {journal} {\bibinfo  {journal} {J.
  Opt. Soc. Am. A}\ }\textbf {\bibinfo {volume} {9}},\ \bibinfo {pages} {2159}
  (\bibinfo {year} {1992})}\BibitemShut {NoStop}%
\bibitem [{\citenamefont {Tey}\ \emph {et~al.}(2008)\citenamefont {Tey},
  \citenamefont {Chen}, \citenamefont {Aljunid}, \citenamefont {Chng},
  \citenamefont {Huber}, \citenamefont {Maslennikov},\ and\ \citenamefont
  {Kurtsiefer}}]{tey2008-np}%
  \BibitemOpen
  \bibfield  {author} {\bibinfo {author} {\bibfnamefont {M.~K.}\ \bibnamefont
  {Tey}}, \bibinfo {author} {\bibfnamefont {Z.}~\bibnamefont {Chen}}, \bibinfo
  {author} {\bibfnamefont {S.~A.}\ \bibnamefont {Aljunid}}, \bibinfo {author}
  {\bibfnamefont {B.}~\bibnamefont {Chng}}, \bibinfo {author} {\bibfnamefont
  {F.}~\bibnamefont {Huber}}, \bibinfo {author} {\bibfnamefont
  {G.}~\bibnamefont {Maslennikov}}, \ and\ \bibinfo {author} {\bibfnamefont
  {C.}~\bibnamefont {Kurtsiefer}},\ }\href {\doibase 10.1038/nphys1096}
  {\bibfield  {journal} {\bibinfo  {journal} {Nature Physics}\ }\textbf
  {\bibinfo {volume} {4}},\ \bibinfo {pages} {924} (\bibinfo {year}
  {2008})}\BibitemShut {NoStop}%
\bibitem [{\citenamefont {Piro}\ \emph {et~al.}(2011)\citenamefont {Piro},
  \citenamefont {Rohde}, \citenamefont {Schuck}, \citenamefont {Almendros},
  \citenamefont {Huwer}, \citenamefont {Ghosh}, \citenamefont {Haase},
  \citenamefont {Hennrich}, \citenamefont {Dubin},\ and\ \citenamefont
  {Eschner}}]{piro2011}%
  \BibitemOpen
  \bibfield  {author} {\bibinfo {author} {\bibfnamefont {N.}~\bibnamefont
  {Piro}}, \bibinfo {author} {\bibfnamefont {F.}~\bibnamefont {Rohde}},
  \bibinfo {author} {\bibfnamefont {C.}~\bibnamefont {Schuck}}, \bibinfo
  {author} {\bibfnamefont {M.}~\bibnamefont {Almendros}}, \bibinfo {author}
  {\bibfnamefont {J.}~\bibnamefont {Huwer}}, \bibinfo {author} {\bibfnamefont
  {J.}~\bibnamefont {Ghosh}}, \bibinfo {author} {\bibfnamefont
  {A.}~\bibnamefont {Haase}}, \bibinfo {author} {\bibfnamefont
  {M.}~\bibnamefont {Hennrich}}, \bibinfo {author} {\bibfnamefont
  {F.}~\bibnamefont {Dubin}}, \ and\ \bibinfo {author} {\bibfnamefont
  {J.}~\bibnamefont {Eschner}},\ }\href {\doibase 10.1038/nphys1805} {\bibfield
   {journal} {\bibinfo  {journal} {Nat. Phys.}\ }\textbf {\bibinfo {volume}
  {7}},\ \bibinfo {pages} {17} (\bibinfo {year} {2011})}\BibitemShut {NoStop}%
\bibitem [{\citenamefont {Aljunid}\ \emph {et~al.}(2013)\citenamefont
  {Aljunid}, \citenamefont {Maslennikov}, \citenamefont {Wang}, \citenamefont
  {Dao}, \citenamefont {Scarani},\ and\ \citenamefont
  {Kurtsiefer}}]{aljunid2013}%
  \BibitemOpen
  \bibfield  {author} {\bibinfo {author} {\bibfnamefont {S.~A.}\ \bibnamefont
  {Aljunid}}, \bibinfo {author} {\bibfnamefont {G.}~\bibnamefont
  {Maslennikov}}, \bibinfo {author} {\bibfnamefont {Y.}~\bibnamefont {Wang}},
  \bibinfo {author} {\bibfnamefont {H.~L.}\ \bibnamefont {Dao}}, \bibinfo
  {author} {\bibfnamefont {V.}~\bibnamefont {Scarani}}, \ and\ \bibinfo
  {author} {\bibfnamefont {C.}~\bibnamefont {Kurtsiefer}},\ }\href {\doibase
  10.1103/PhysRevLett.111.103001} {\bibfield  {journal} {\bibinfo  {journal}
  {Phys. Rev. Lett.}\ }\textbf {\bibinfo {volume} {111}},\ \bibinfo {pages}
  {103001} (\bibinfo {year} {2013})},\ \Eprint
  {http://arxiv.org/abs/arXiv:1304.3761} {arXiv:1304.3761} \BibitemShut
  {NoStop}%
\bibitem [{\citenamefont {Gardiner}(1986)}]{gardiner1986}%
  \BibitemOpen
  \bibfield  {author} {\bibinfo {author} {\bibfnamefont {C.~W.}\ \bibnamefont
  {Gardiner}},\ }\href {\doibase 10.1103/PhysRevLett.56.1917} {\bibfield
  {journal} {\bibinfo  {journal} {Phys. Rev. Lett.}\ }\textbf {\bibinfo
  {volume} {56}},\ \bibinfo {pages} {1917} (\bibinfo {year}
  {1986})}\BibitemShut {NoStop}%
\bibitem [{\citenamefont {Bokor}\ and\ \citenamefont
  {Davidson}(2008)}]{bokor2008}%
  \BibitemOpen
  \bibfield  {author} {\bibinfo {author} {\bibfnamefont {N.}~\bibnamefont
  {Bokor}}\ and\ \bibinfo {author} {\bibfnamefont {N.}~\bibnamefont
  {Davidson}},\ }\href@noop {} {\bibfield  {journal} {\bibinfo  {journal} {Opt.
  Commun.}\ }\textbf {\bibinfo {volume} {281}},\ \bibinfo {pages} {5499}
  (\bibinfo {year} {2008})}\BibitemShut {NoStop}%
\bibitem [{\citenamefont {Leuchs}\ \emph {et~al.}(2008)\citenamefont {Leuchs},
  \citenamefont {Mantel}, \citenamefont {Berger}, \citenamefont {Konermann},
  \citenamefont {Sondermann}, \citenamefont {Peschel}, \citenamefont
  {Lindlein},\ and\ \citenamefont {Schwider}}]{leuchs2008}%
  \BibitemOpen
  \bibfield  {author} {\bibinfo {author} {\bibfnamefont {G.}~\bibnamefont
  {Leuchs}}, \bibinfo {author} {\bibfnamefont {K.}~\bibnamefont {Mantel}},
  \bibinfo {author} {\bibfnamefont {A.}~\bibnamefont {Berger}}, \bibinfo
  {author} {\bibfnamefont {H.}~\bibnamefont {Konermann}}, \bibinfo {author}
  {\bibfnamefont {M.}~\bibnamefont {Sondermann}}, \bibinfo {author}
  {\bibfnamefont {U.}~\bibnamefont {Peschel}}, \bibinfo {author} {\bibfnamefont
  {N.}~\bibnamefont {Lindlein}}, \ and\ \bibinfo {author} {\bibfnamefont
  {J.}~\bibnamefont {Schwider}},\ }\href {\doibase 10.1364/AO.47.005570}
  {\bibfield  {journal} {\bibinfo  {journal} {Applied Optics}\ }\textbf
  {\bibinfo {volume} {47}},\ \bibinfo {pages} {5570} (\bibinfo {year}
  {2008})}\BibitemShut {NoStop}%
\bibitem [{\citenamefont {Maiwald}\ \emph {et~al.}(2012)\citenamefont
  {Maiwald}, \citenamefont {Golla}, \citenamefont {Fischer}, \citenamefont
  {Bader}, \citenamefont {Heugel}, \citenamefont {Chalopin}, \citenamefont
  {Sondermann},\ and\ \citenamefont {Leuchs}}]{maiwald2012}%
  \BibitemOpen
  \bibfield  {author} {\bibinfo {author} {\bibfnamefont {R.}~\bibnamefont
  {Maiwald}}, \bibinfo {author} {\bibfnamefont {A.}~\bibnamefont {Golla}},
  \bibinfo {author} {\bibfnamefont {M.}~\bibnamefont {Fischer}}, \bibinfo
  {author} {\bibfnamefont {M.}~\bibnamefont {Bader}}, \bibinfo {author}
  {\bibfnamefont {S.}~\bibnamefont {Heugel}}, \bibinfo {author} {\bibfnamefont
  {B.}~\bibnamefont {Chalopin}}, \bibinfo {author} {\bibfnamefont
  {M.}~\bibnamefont {Sondermann}}, \ and\ \bibinfo {author} {\bibfnamefont
  {G.}~\bibnamefont {Leuchs}},\ }\href {\doibase 10.1103/PhysRevA.86.043431}
  {\bibfield  {journal} {\bibinfo  {journal} {Phys. Rev. A}\ }\textbf {\bibinfo
  {volume} {86}},\ \bibinfo {pages} {043431} (\bibinfo {year}
  {2012})}\BibitemShut {NoStop}%
\bibitem [{\citenamefont {Slodi\ifmmode~\check{c}\else \v{c}\fi{}ka}\ \emph
  {et~al.}(2010)\citenamefont {Slodi\ifmmode~\check{c}\else \v{c}\fi{}ka},
  \citenamefont {H\'etet}, \citenamefont {Gerber}, \citenamefont {Hennrich},\
  and\ \citenamefont {Blatt}}]{slodicka2010}%
  \BibitemOpen
  \bibfield  {author} {\bibinfo {author} {\bibfnamefont {L.}~\bibnamefont
  {Slodi\ifmmode~\check{c}\else \v{c}\fi{}ka}}, \bibinfo {author}
  {\bibfnamefont {G.}~\bibnamefont {H\'etet}}, \bibinfo {author} {\bibfnamefont
  {S.}~\bibnamefont {Gerber}}, \bibinfo {author} {\bibfnamefont
  {M.}~\bibnamefont {Hennrich}}, \ and\ \bibinfo {author} {\bibfnamefont
  {R.}~\bibnamefont {Blatt}},\ }\href {\doibase 10.1103/PhysRevLett.105.153604}
  {\bibfield  {journal} {\bibinfo  {journal} {Phys. Rev. Lett.}\ }\textbf
  {\bibinfo {volume} {105}},\ \bibinfo {pages} {153604} (\bibinfo {year}
  {2010})}\BibitemShut {NoStop}%
\bibitem [{\citenamefont {Bomzon}\ \emph {et~al.}(2002)\citenamefont {Bomzon},
  \citenamefont {Biener}, \citenamefont {Kleiner},\ and\ \citenamefont
  {Hasman}}]{bomzon2002}%
  \BibitemOpen
  \bibfield  {author} {\bibinfo {author} {\bibfnamefont {Z.}~\bibnamefont
  {Bomzon}}, \bibinfo {author} {\bibfnamefont {G.}~\bibnamefont {Biener}},
  \bibinfo {author} {\bibfnamefont {V.}~\bibnamefont {Kleiner}}, \ and\
  \bibinfo {author} {\bibfnamefont {E.}~\bibnamefont {Hasman}},\ }\href
  {\doibase 10.1364/OL.27.000285} {\bibfield  {journal} {\bibinfo  {journal}
  {Opt. Lett.}\ }\textbf {\bibinfo {volume} {27}},\ \bibinfo {pages} {285}
  (\bibinfo {year} {2002})}\BibitemShut {NoStop}%
\bibitem [{\citenamefont {Ghadyani}\ \emph {et~al.}(2011)\citenamefont
  {Ghadyani}, \citenamefont {Vartiainen}, \citenamefont {Harder}, \citenamefont
  {Iff}, \citenamefont {Berger}, \citenamefont {Lindlein},\ and\ \citenamefont
  {Kuittinen}}]{ghadyani2011}%
  \BibitemOpen
  \bibfield  {author} {\bibinfo {author} {\bibfnamefont {Z.}~\bibnamefont
  {Ghadyani}}, \bibinfo {author} {\bibfnamefont {I.}~\bibnamefont
  {Vartiainen}}, \bibinfo {author} {\bibfnamefont {I.}~\bibnamefont {Harder}},
  \bibinfo {author} {\bibfnamefont {W.}~\bibnamefont {Iff}}, \bibinfo {author}
  {\bibfnamefont {A.}~\bibnamefont {Berger}}, \bibinfo {author} {\bibfnamefont
  {N.}~\bibnamefont {Lindlein}}, \ and\ \bibinfo {author} {\bibfnamefont
  {M.}~\bibnamefont {Kuittinen}},\ }\href {\doibase 10.1364/AO.50.002451}
  {\bibfield  {journal} {\bibinfo  {journal} {Appl. Opt.}\ }\textbf {\bibinfo
  {volume} {50}},\ \bibinfo {pages} {2451} (\bibinfo {year}
  {2011})}\BibitemShut {NoStop}%
\bibitem [{\citenamefont {Stalder}\ and\ \citenamefont
  {Schadt}(1996)}]{stalder1996}%
  \BibitemOpen
  \bibfield  {author} {\bibinfo {author} {\bibfnamefont {M.}~\bibnamefont
  {Stalder}}\ and\ \bibinfo {author} {\bibfnamefont {M.}~\bibnamefont
  {Schadt}},\ }\href@noop {} {\bibfield  {journal} {\bibinfo  {journal} {Opt.
  Lett.}\ }\textbf {\bibinfo {volume} {21}},\ \bibinfo {pages} {1948} (\bibinfo
  {year} {1996})}\BibitemShut {NoStop}%
\bibitem [{\citenamefont {Tidwell}\ \emph {et~al.}(1990)\citenamefont
  {Tidwell}, \citenamefont {Ford},\ and\ \citenamefont {Kimura}}]{tidwell1990}%
  \BibitemOpen
  \bibfield  {author} {\bibinfo {author} {\bibfnamefont {S.}~\bibnamefont
  {Tidwell}}, \bibinfo {author} {\bibfnamefont {D.}~\bibnamefont {Ford}}, \
  and\ \bibinfo {author} {\bibfnamefont {W.}~\bibnamefont {Kimura}},\
  }\href@noop {} {\bibfield  {journal} {\bibinfo  {journal} {Applied Optics}\
  }\textbf {\bibinfo {volume} {29}},\ \bibinfo {pages} {2234} (\bibinfo {year}
  {1990})}\BibitemShut {NoStop}%
\bibitem [{\citenamefont {Tidwell}\ \emph {et~al.}(1993)\citenamefont
  {Tidwell}, \citenamefont {Kim},\ and\ \citenamefont {Kimura}}]{tidwell1993}%
  \BibitemOpen
  \bibfield  {author} {\bibinfo {author} {\bibfnamefont {S.}~\bibnamefont
  {Tidwell}}, \bibinfo {author} {\bibfnamefont {G.}~\bibnamefont {Kim}}, \ and\
  \bibinfo {author} {\bibfnamefont {W.}~\bibnamefont {Kimura}},\ }\href@noop {}
  {\bibfield  {journal} {\bibinfo  {journal} {Applied Optics}\ }\textbf
  {\bibinfo {volume} {32}},\ \bibinfo {pages} {5222} (\bibinfo {year}
  {1993})}\BibitemShut {NoStop}%
\bibitem [{\citenamefont {Oron}\ \emph {et~al.}(2000)\citenamefont {Oron},
  \citenamefont {Blit}, \citenamefont {Davidson}, \citenamefont {Friesem},
  \citenamefont {Bomzon},\ and\ \citenamefont {Hasman}}]{oron2000}%
  \BibitemOpen
  \bibfield  {author} {\bibinfo {author} {\bibfnamefont {R.}~\bibnamefont
  {Oron}}, \bibinfo {author} {\bibfnamefont {S.}~\bibnamefont {Blit}}, \bibinfo
  {author} {\bibfnamefont {N.}~\bibnamefont {Davidson}}, \bibinfo {author}
  {\bibfnamefont {A.}~\bibnamefont {Friesem}}, \bibinfo {author} {\bibfnamefont
  {Z.}~\bibnamefont {Bomzon}}, \ and\ \bibinfo {author} {\bibfnamefont
  {E.}~\bibnamefont {Hasman}},\ }\href@noop {} {\bibfield  {journal} {\bibinfo
  {journal} {Applied Physics Letters}\ }\textbf {\bibinfo {volume} {77}},\
  \bibinfo {pages} {3322} (\bibinfo {year} {2000})}\BibitemShut {NoStop}%
\bibitem [{\citenamefont {Maurer}\ \emph {et~al.}(2007)\citenamefont {Maurer},
  \citenamefont {Jesacher}, \citenamefont {F\"urhapter}, \citenamefont
  {Bernet},\ and\ \citenamefont {Ritsch-Marte}}]{maurer2007}%
  \BibitemOpen
  \bibfield  {author} {\bibinfo {author} {\bibfnamefont {C.}~\bibnamefont
  {Maurer}}, \bibinfo {author} {\bibfnamefont {A.}~\bibnamefont {Jesacher}},
  \bibinfo {author} {\bibfnamefont {S.}~\bibnamefont {F\"urhapter}}, \bibinfo
  {author} {\bibfnamefont {S.}~\bibnamefont {Bernet}}, \ and\ \bibinfo {author}
  {\bibfnamefont {M.}~\bibnamefont {Ritsch-Marte}},\ }\href {\doibase
  10.1088/1367-2630/9/3/078} {\bibfield  {journal} {\bibinfo  {journal} {New
  Journal of Physics}\ }\textbf {\bibinfo {volume} {9}},\ \bibinfo {pages} {78}
  (\bibinfo {year} {2007})}\BibitemShut {NoStop}%
\bibitem [{\citenamefont {Lin}\ \emph {et~al.}(2013)\citenamefont {Lin},
  \citenamefont {Genevet}, \citenamefont {Kats}, \citenamefont {Antoniou},\
  and\ \citenamefont {Capasso}}]{lin2013}%
  \BibitemOpen
  \bibfield  {author} {\bibinfo {author} {\bibfnamefont {J.}~\bibnamefont
  {Lin}}, \bibinfo {author} {\bibfnamefont {P.}~\bibnamefont {Genevet}},
  \bibinfo {author} {\bibfnamefont {M.~A.}\ \bibnamefont {Kats}}, \bibinfo
  {author} {\bibfnamefont {N.}~\bibnamefont {Antoniou}}, \ and\ \bibinfo
  {author} {\bibfnamefont {F.}~\bibnamefont {Capasso}},\ }\href@noop {}
  {\bibfield  {journal} {\bibinfo  {journal} {Nano letters}\ }\textbf {\bibinfo
  {volume} {13}},\ \bibinfo {pages} {4269} (\bibinfo {year}
  {2013})}\BibitemShut {NoStop}%
\bibitem [{\citenamefont {Dorn}\ \emph {et~al.}(2003)\citenamefont {Dorn},
  \citenamefont {Quabis},\ and\ \citenamefont {Leuchs}}]{dorn2003}%
  \BibitemOpen
  \bibfield  {author} {\bibinfo {author} {\bibfnamefont {R.}~\bibnamefont
  {Dorn}}, \bibinfo {author} {\bibfnamefont {S.}~\bibnamefont {Quabis}}, \ and\
  \bibinfo {author} {\bibfnamefont {G.}~\bibnamefont {Leuchs}},\ }\href@noop {}
  {\bibfield  {journal} {\bibinfo  {journal} {Phys. Rev. Lett.}\ }\textbf
  {\bibinfo {volume} {91}},\ \bibinfo {pages} {233901} (\bibinfo {year}
  {2003})}\BibitemShut {NoStop}%
\bibitem [{\citenamefont {Quabis}\ \emph {et~al.}(2005)\citenamefont {Quabis},
  \citenamefont {Dorn},\ and\ \citenamefont {Leuchs}}]{quabis2005}%
  \BibitemOpen
  \bibfield  {author} {\bibinfo {author} {\bibfnamefont {S.}~\bibnamefont
  {Quabis}}, \bibinfo {author} {\bibfnamefont {R.}~\bibnamefont {Dorn}}, \ and\
  \bibinfo {author} {\bibfnamefont {G.}~\bibnamefont {Leuchs}},\ }\href@noop {}
  {\bibfield  {journal} {\bibinfo  {journal} {Appl. Phys. B}\ }\textbf
  {\bibinfo {volume} {81}},\ \bibinfo {pages} {597} (\bibinfo {year}
  {2005})}\BibitemShut {NoStop}%
\bibitem [{\citenamefont {Born}\ and\ \citenamefont
  {Wolf}(1991)}]{born-wolf1991}%
  \BibitemOpen
  \bibfield  {author} {\bibinfo {author} {\bibfnamefont {M.}~\bibnamefont
  {Born}}\ and\ \bibinfo {author} {\bibfnamefont {E.}~\bibnamefont {Wolf}},\
  }\href@noop {} {\emph {\bibinfo {title} {Principles of optics}}},\ \bibinfo
  {edition} {6th}\ ed.\ (\bibinfo  {publisher} {Pergamon Press},\ \bibinfo
  {address} {Oxford},\ \bibinfo {year} {1991})\BibitemShut {NoStop}%
\bibitem [{\citenamefont {Aljunid}\ \emph {et~al.}(2009)\citenamefont
  {Aljunid}, \citenamefont {Tey}, \citenamefont {Chng}, \citenamefont {Liew},
  \citenamefont {Maslennikov}, \citenamefont {Scarani},\ and\ \citenamefont
  {Kurtsiefer}}]{aljunid2009}%
  \BibitemOpen
  \bibfield  {author} {\bibinfo {author} {\bibfnamefont {S.~A.}\ \bibnamefont
  {Aljunid}}, \bibinfo {author} {\bibfnamefont {M.~K.}\ \bibnamefont {Tey}},
  \bibinfo {author} {\bibfnamefont {B.}~\bibnamefont {Chng}}, \bibinfo {author}
  {\bibfnamefont {T.}~\bibnamefont {Liew}}, \bibinfo {author} {\bibfnamefont
  {G.}~\bibnamefont {Maslennikov}}, \bibinfo {author} {\bibfnamefont
  {V.}~\bibnamefont {Scarani}}, \ and\ \bibinfo {author} {\bibfnamefont
  {C.}~\bibnamefont {Kurtsiefer}},\ }\href {\doibase
  10.1103/PhysRevLett.103.153601} {\bibfield  {journal} {\bibinfo  {journal}
  {Physical Review Letters}\ }\textbf {\bibinfo {volume} {103}},\ \bibinfo
  {eid} {153601} (\bibinfo {year} {2009})}\BibitemShut {NoStop}%
\bibitem [{\citenamefont {Pototschnig}\ \emph {et~al.}(2011)\citenamefont
  {Pototschnig}, \citenamefont {Chassagneux}, \citenamefont {Hwang},
  \citenamefont {Zumofen}, \citenamefont {Renn},\ and\ \citenamefont
  {Sandoghdar}}]{pototschnig2011}%
  \BibitemOpen
  \bibfield  {author} {\bibinfo {author} {\bibfnamefont {M.}~\bibnamefont
  {Pototschnig}}, \bibinfo {author} {\bibfnamefont {Y.}~\bibnamefont
  {Chassagneux}}, \bibinfo {author} {\bibfnamefont {J.}~\bibnamefont {Hwang}},
  \bibinfo {author} {\bibfnamefont {G.}~\bibnamefont {Zumofen}}, \bibinfo
  {author} {\bibfnamefont {A.}~\bibnamefont {Renn}}, \ and\ \bibinfo {author}
  {\bibfnamefont {V.}~\bibnamefont {Sandoghdar}},\ }\href {\doibase
  10.1103/PhysRevLett.107.063001} {\bibfield  {journal} {\bibinfo  {journal}
  {Phys. Rev. Lett.}\ }\textbf {\bibinfo {volume} {107}},\ \bibinfo {pages}
  {063001} (\bibinfo {year} {2011})}\BibitemShut {NoStop}%
\bibitem [{\citenamefont {H\'etet}\ \emph {et~al.}(2013)\citenamefont
  {H\'etet}, \citenamefont {Slodi\ifmmode~\check{c}\else \v{c}\fi{}ka},
  \citenamefont {R\"ock},\ and\ \citenamefont {Blatt}}]{hetet2013}%
  \BibitemOpen
  \bibfield  {author} {\bibinfo {author} {\bibfnamefont {G.}~\bibnamefont
  {H\'etet}}, \bibinfo {author} {\bibfnamefont {L.}~\bibnamefont
  {Slodi\ifmmode~\check{c}\else \v{c}\fi{}ka}}, \bibinfo {author}
  {\bibfnamefont {N.}~\bibnamefont {R\"ock}}, \ and\ \bibinfo {author}
  {\bibfnamefont {R.}~\bibnamefont {Blatt}},\ }\href {\doibase
  10.1103/PhysRevA.88.041804} {\bibfield  {journal} {\bibinfo  {journal} {Phys.
  Rev. A}\ }\textbf {\bibinfo {volume} {88}},\ \bibinfo {pages} {041804}
  (\bibinfo {year} {2013})}\BibitemShut {NoStop}%
\bibitem [{\citenamefont {Jechow}\ \emph {et~al.}(2013)\citenamefont {Jechow},
  \citenamefont {Norton}, \citenamefont {H\"andel}, \citenamefont
  {Bl\ifmmode~\bar{u}\else \={u}\fi{}ms}, \citenamefont {Streed},\ and\
  \citenamefont {Kielpinski}}]{jechow2013}%
  \BibitemOpen
  \bibfield  {author} {\bibinfo {author} {\bibfnamefont {A.}~\bibnamefont
  {Jechow}}, \bibinfo {author} {\bibfnamefont {B.~G.}\ \bibnamefont {Norton}},
  \bibinfo {author} {\bibfnamefont {S.}~\bibnamefont {H\"andel}}, \bibinfo
  {author} {\bibfnamefont {V.}~\bibnamefont {Bl\ifmmode~\bar{u}\else
  \={u}\fi{}ms}}, \bibinfo {author} {\bibfnamefont {E.~W.}\ \bibnamefont
  {Streed}}, \ and\ \bibinfo {author} {\bibfnamefont {D.}~\bibnamefont
  {Kielpinski}},\ }\href {\doibase 10.1103/PhysRevLett.110.113605} {\bibfield
  {journal} {\bibinfo  {journal} {Phys. Rev. Lett.}\ }\textbf {\bibinfo
  {volume} {110}},\ \bibinfo {pages} {113605} (\bibinfo {year} {2013})},\
  \Eprint {http://arxiv.org/abs/arXiv:1208.5091} {arXiv:1208.5091} \BibitemShut
  {NoStop}%
\bibitem [{\citenamefont {Zumofen}\ \emph {et~al.}(2008)\citenamefont
  {Zumofen}, \citenamefont {Mojarad}, \citenamefont {Sandoghdar},\ and\
  \citenamefont {Agio}}]{zumofen2008}%
  \BibitemOpen
  \bibfield  {author} {\bibinfo {author} {\bibfnamefont {G.}~\bibnamefont
  {Zumofen}}, \bibinfo {author} {\bibfnamefont {N.~M.}\ \bibnamefont
  {Mojarad}}, \bibinfo {author} {\bibfnamefont {V.}~\bibnamefont {Sandoghdar}},
  \ and\ \bibinfo {author} {\bibfnamefont {M.}~\bibnamefont {Agio}},\ }\href
  {\doibase 10.1103/PhysRevLett.101.180404} {\bibfield  {journal} {\bibinfo
  {journal} {Phys. Rev. Lett.}\ }\textbf {\bibinfo {volume} {101}},\ \bibinfo
  {pages} {180404} (\bibinfo {year} {2008})}\BibitemShut {NoStop}%
\bibitem [{\citenamefont {Tey}\ \emph {et~al.}(2009)\citenamefont {Tey},
  \citenamefont {Maslennikov}, \citenamefont {Liew}, \citenamefont {Aljunid},
  \citenamefont {Huber}, \citenamefont {Chng}, \citenamefont {Chen},
  \citenamefont {Scarani},\ and\ \citenamefont {Kurtsiefer}}]{tey2009}%
  \BibitemOpen
  \bibfield  {author} {\bibinfo {author} {\bibfnamefont {M.~K.}\ \bibnamefont
  {Tey}}, \bibinfo {author} {\bibfnamefont {G.}~\bibnamefont {Maslennikov}},
  \bibinfo {author} {\bibfnamefont {T.~C.~H.}\ \bibnamefont {Liew}}, \bibinfo
  {author} {\bibfnamefont {S.~A.}\ \bibnamefont {Aljunid}}, \bibinfo {author}
  {\bibfnamefont {F.}~\bibnamefont {Huber}}, \bibinfo {author} {\bibfnamefont
  {B.}~\bibnamefont {Chng}}, \bibinfo {author} {\bibfnamefont {Z.}~\bibnamefont
  {Chen}}, \bibinfo {author} {\bibfnamefont {V.}~\bibnamefont {Scarani}}, \
  and\ \bibinfo {author} {\bibfnamefont {C.}~\bibnamefont {Kurtsiefer}},\
  }\href@noop {} {\bibfield  {journal} {\bibinfo  {journal} {New Journal of
  Physics}\ }\textbf {\bibinfo {volume} {11}},\ \bibinfo {pages} {043011}
  (\bibinfo {year} {2009})}\BibitemShut {NoStop}%
\bibitem [{\citenamefont {Tyc}(2012)}]{tyc2012}%
  \BibitemOpen
  \bibfield  {author} {\bibinfo {author} {\bibfnamefont {T.}~\bibnamefont
  {Tyc}},\ }\href {\doibase 10.1364/OL.37.000924} {\bibfield  {journal}
  {\bibinfo  {journal} {Opt. Lett.}\ }\textbf {\bibinfo {volume} {37}},\
  \bibinfo {pages} {924} (\bibinfo {year} {2012})}\BibitemShut {NoStop}%
\bibitem [{\citenamefont {Sondermann}\ and\ \citenamefont
  {Leuchs}(2013{\natexlab{a}})}]{sondermann2013p}%
  \BibitemOpen
  \bibfield  {author} {\bibinfo {author} {\bibfnamefont {M.}~\bibnamefont
  {Sondermann}}\ and\ \bibinfo {author} {\bibfnamefont {G.}~\bibnamefont
  {Leuchs}},\ }\href {\doibase 10.2971/jeos.2013.13052} {\bibfield  {journal}
  {\bibinfo  {journal} {J. Europ. Opt. Soc. Rap. Public.}\ }\textbf {\bibinfo
  {volume} {8}},\ \bibinfo {pages} {13502} (\bibinfo {year}
  {2013}{\natexlab{a}})},\ \Eprint {http://arxiv.org/abs/arXiv:1306.2804
  [quant-ph]} {arXiv:1306.2804 [quant-ph]} \BibitemShut {NoStop}%
\bibitem [{\citenamefont {Vamivakas}\ \emph {et~al.}(2007)\citenamefont
  {Vamivakas}, \citenamefont {Atat\"ure}, \citenamefont {Dreiser},
  \citenamefont {Yilmaz}, \citenamefont {Badolato}, \citenamefont {Swan},
  \citenamefont {Goldberg}, \citenamefont {Imamoglu},\ and\ \citenamefont
  {\"Unl\"u}}]{vamivakas2007}%
  \BibitemOpen
  \bibfield  {author} {\bibinfo {author} {\bibfnamefont {A.~N.}\ \bibnamefont
  {Vamivakas}}, \bibinfo {author} {\bibfnamefont {M.}~\bibnamefont
  {Atat\"ure}}, \bibinfo {author} {\bibfnamefont {J.}~\bibnamefont {Dreiser}},
  \bibinfo {author} {\bibfnamefont {S.~T.}\ \bibnamefont {Yilmaz}}, \bibinfo
  {author} {\bibfnamefont {A.}~\bibnamefont {Badolato}}, \bibinfo {author}
  {\bibfnamefont {A.~K.}\ \bibnamefont {Swan}}, \bibinfo {author}
  {\bibfnamefont {B.~B.}\ \bibnamefont {Goldberg}}, \bibinfo {author}
  {\bibfnamefont {A.}~\bibnamefont {Imamoglu}}, \ and\ \bibinfo {author}
  {\bibfnamefont {M.~S.}\ \bibnamefont {\"Unl\"u}},\ }\href@noop {} {\bibfield
  {journal} {\bibinfo  {journal} {Nano Letters}\ }\textbf {\bibinfo {volume}
  {7}},\ \bibinfo {pages} {2892} (\bibinfo {year} {2007})}\BibitemShut
  {NoStop}%
\bibitem [{\citenamefont {Wrigge}\ \emph {et~al.}(2008)\citenamefont {Wrigge},
  \citenamefont {Gerhardt}, \citenamefont {Hwang}, \citenamefont {Zumofen},\
  and\ \citenamefont {Sandoghdar}}]{wrigge2008}%
  \BibitemOpen
  \bibfield  {author} {\bibinfo {author} {\bibfnamefont {G.}~\bibnamefont
  {Wrigge}}, \bibinfo {author} {\bibfnamefont {I.}~\bibnamefont {Gerhardt}},
  \bibinfo {author} {\bibfnamefont {J.}~\bibnamefont {Hwang}}, \bibinfo
  {author} {\bibfnamefont {G.}~\bibnamefont {Zumofen}}, \ and\ \bibinfo
  {author} {\bibfnamefont {V.}~\bibnamefont {Sandoghdar}},\ }\href@noop {}
  {\bibfield  {journal} {\bibinfo  {journal} {Nature Physics}\ }\textbf
  {\bibinfo {volume} {4}},\ \bibinfo {pages} {60} (\bibinfo {year}
  {2008})}\BibitemShut {NoStop}%
\bibitem [{\citenamefont {Aljunid}\ \emph {et~al.}(2011)\citenamefont
  {Aljunid}, \citenamefont {Chng}, \citenamefont {Lee}, \citenamefont
  {Paesold}, \citenamefont {Maslennikov},\ and\ \citenamefont
  {Kurtsiefer}}]{aljunid2011}%
  \BibitemOpen
  \bibfield  {author} {\bibinfo {author} {\bibfnamefont {S.~A.}\ \bibnamefont
  {Aljunid}}, \bibinfo {author} {\bibfnamefont {B.}~\bibnamefont {Chng}},
  \bibinfo {author} {\bibfnamefont {J.}~\bibnamefont {Lee}}, \bibinfo {author}
  {\bibfnamefont {M.}~\bibnamefont {Paesold}}, \bibinfo {author} {\bibfnamefont
  {G.}~\bibnamefont {Maslennikov}}, \ and\ \bibinfo {author} {\bibfnamefont
  {C.}~\bibnamefont {Kurtsiefer}},\ }\href {\doibase
  10.1080/09500340.2010.522780} {\bibfield  {journal} {\bibinfo  {journal}
  {Journal of Modern Optics}\ }\textbf {\bibinfo {volume} {58}},\ \bibinfo
  {pages} {299} (\bibinfo {year} {2011})}\BibitemShut {NoStop}%
\bibitem [{\citenamefont {Rezus}\ \emph {et~al.}(2012)\citenamefont {Rezus},
  \citenamefont {Walt}, \citenamefont {Lettow}, \citenamefont {Renn},
  \citenamefont {Zumofen}, \citenamefont {G\"otzinger},\ and\ \citenamefont
  {Sandoghdar}}]{rezus2012}%
  \BibitemOpen
  \bibfield  {author} {\bibinfo {author} {\bibfnamefont {Y.~L.~A.}\
  \bibnamefont {Rezus}}, \bibinfo {author} {\bibfnamefont {S.~G.}\ \bibnamefont
  {Walt}}, \bibinfo {author} {\bibfnamefont {R.}~\bibnamefont {Lettow}},
  \bibinfo {author} {\bibfnamefont {A.}~\bibnamefont {Renn}}, \bibinfo {author}
  {\bibfnamefont {G.}~\bibnamefont {Zumofen}}, \bibinfo {author} {\bibfnamefont
  {S.}~\bibnamefont {G\"otzinger}}, \ and\ \bibinfo {author} {\bibfnamefont
  {V.}~\bibnamefont {Sandoghdar}},\ }\href {\doibase
  10.1103/PhysRevLett.108.093601} {\bibfield  {journal} {\bibinfo  {journal}
  {Phys. Rev. Lett.}\ }\textbf {\bibinfo {volume} {108}},\ \bibinfo {pages}
  {093601} (\bibinfo {year} {2012})}\BibitemShut {NoStop}%
\bibitem [{\citenamefont {Kochan}\ and\ \citenamefont
  {Carmichael}(1994)}]{kochan1994}%
  \BibitemOpen
  \bibfield  {author} {\bibinfo {author} {\bibfnamefont {P.}~\bibnamefont
  {Kochan}}\ and\ \bibinfo {author} {\bibfnamefont {H.~J.}\ \bibnamefont
  {Carmichael}},\ }\href {\doibase 10.1103/PhysRevA.50.1700} {\bibfield
  {journal} {\bibinfo  {journal} {Phys. Rev. A}\ }\textbf {\bibinfo {volume}
  {50}},\ \bibinfo {pages} {1700} (\bibinfo {year} {1994})}\BibitemShut
  {NoStop}%
\bibitem [{\citenamefont {H\'etet}\ \emph {et~al.}(2011)\citenamefont
  {H\'etet}, \citenamefont {Slodi\ifmmode~\check{c}\else \v{c}\fi{}ka},
  \citenamefont {Hennrich},\ and\ \citenamefont {Blatt}}]{hetet2011}%
  \BibitemOpen
  \bibfield  {author} {\bibinfo {author} {\bibfnamefont {G.}~\bibnamefont
  {H\'etet}}, \bibinfo {author} {\bibfnamefont {L.}~\bibnamefont
  {Slodi\ifmmode~\check{c}\else \v{c}\fi{}ka}}, \bibinfo {author}
  {\bibfnamefont {M.}~\bibnamefont {Hennrich}}, \ and\ \bibinfo {author}
  {\bibfnamefont {R.}~\bibnamefont {Blatt}},\ }\href {\doibase
  10.1103/PhysRevLett.107.133002} {\bibfield  {journal} {\bibinfo  {journal}
  {Phys. Rev. Lett.}\ }\textbf {\bibinfo {volume} {107}},\ \bibinfo {pages}
  {133002} (\bibinfo {year} {2011})}\BibitemShut {NoStop}%
\bibitem [{\citenamefont {Sondermann}\ and\ \citenamefont
  {Leuchs}(2013{\natexlab{b}})}]{sondermann2013s}%
  \BibitemOpen
  \bibfield  {author} {\bibinfo {author} {\bibfnamefont {M.}~\bibnamefont
  {Sondermann}}\ and\ \bibinfo {author} {\bibfnamefont {G.}~\bibnamefont
  {Leuchs}},\ }\href@noop {} {\bibfield  {journal} {\bibinfo  {journal}
  {Romanian Reports in Physics}\ }\textbf {\bibinfo {volume} {65}},\ \bibinfo
  {pages} {638} (\bibinfo {year} {2013}{\natexlab{b}})},\ \Eprint
  {http://arxiv.org/abs/arXiv:1306.3902 [quant-ph]} {arXiv:1306.3902
  [quant-ph]} \BibitemShut {NoStop}%
\bibitem [{\citenamefont {Yamamoto}\ and\ \citenamefont
  {Haus}(1986)}]{yamamoto1986}%
  \BibitemOpen
  \bibfield  {author} {\bibinfo {author} {\bibfnamefont {Y.}~\bibnamefont
  {Yamamoto}}\ and\ \bibinfo {author} {\bibfnamefont {H.~A.}\ \bibnamefont
  {Haus}},\ }\href {\doibase 10.1103/RevModPhys.58.1001} {\bibfield  {journal}
  {\bibinfo  {journal} {Rev. Mod. Phys.}\ }\textbf {\bibinfo {volume} {58}},\
  \bibinfo {pages} {1001} (\bibinfo {year} {1986})}\BibitemShut {NoStop}%
\bibitem [{\citenamefont {Gaeta}\ and\ \citenamefont {Boyd}(1988)}]{gaeta1988}%
  \BibitemOpen
  \bibfield  {author} {\bibinfo {author} {\bibfnamefont {A.~L.}\ \bibnamefont
  {Gaeta}}\ and\ \bibinfo {author} {\bibfnamefont {R.~W.}\ \bibnamefont
  {Boyd}},\ }\href {\doibase 10.1103/PhysRevLett.60.2618} {\bibfield  {journal}
  {\bibinfo  {journal} {Phys. Rev. Lett.}\ }\textbf {\bibinfo {volume} {60}},\
  \bibinfo {pages} {2618} (\bibinfo {year} {1988})}\BibitemShut {NoStop}%
\bibitem [{\citenamefont {Stobinska}\ \emph {et~al.}(2009)\citenamefont
  {Stobinska}, \citenamefont {Alber},\ and\ \citenamefont
  {Leuchs}}]{stobinska2009}%
  \BibitemOpen
  \bibfield  {author} {\bibinfo {author} {\bibfnamefont {M.}~\bibnamefont
  {Stobinska}}, \bibinfo {author} {\bibfnamefont {G.}~\bibnamefont {Alber}}, \
  and\ \bibinfo {author} {\bibfnamefont {G.}~\bibnamefont {Leuchs}},\ }\href
  {\doibase 10.1209/0295-5075/86/14007} {\bibfield  {journal} {\bibinfo
  {journal} {EPL}\ }\textbf {\bibinfo {volume} {86}},\ \bibinfo {pages} {14007}
  (\bibinfo {year} {2009})},\ \Eprint {http://arxiv.org/abs/arxiv:0808.1666}
  {arxiv:0808.1666} \BibitemShut {NoStop}%
\bibitem [{\citenamefont {Schauer}\ \emph {et~al.}(2010)\citenamefont
  {Schauer}, \citenamefont {Danielson}, \citenamefont {Feldbaum}, \citenamefont
  {Rahaman}, \citenamefont {Wang}, \citenamefont {Zhang}, \citenamefont
  {Zhao},\ and\ \citenamefont {Torgerson}}]{schauer2010}%
  \BibitemOpen
  \bibfield  {author} {\bibinfo {author} {\bibfnamefont {M.~M.}\ \bibnamefont
  {Schauer}}, \bibinfo {author} {\bibfnamefont {J.~R.}\ \bibnamefont
  {Danielson}}, \bibinfo {author} {\bibfnamefont {D.}~\bibnamefont {Feldbaum}},
  \bibinfo {author} {\bibfnamefont {M.~S.}\ \bibnamefont {Rahaman}}, \bibinfo
  {author} {\bibfnamefont {L.-B.}\ \bibnamefont {Wang}}, \bibinfo {author}
  {\bibfnamefont {J.}~\bibnamefont {Zhang}}, \bibinfo {author} {\bibfnamefont
  {X.}~\bibnamefont {Zhao}}, \ and\ \bibinfo {author} {\bibfnamefont {J.~R.}\
  \bibnamefont {Torgerson}},\ }\href {\doibase 10.1103/PhysRevA.82.062518}
  {\bibfield  {journal} {\bibinfo  {journal} {Phys. Rev. A}\ }\textbf {\bibinfo
  {volume} {82}},\ \bibinfo {pages} {062518} (\bibinfo {year}
  {2010})}\BibitemShut {NoStop}%
\bibitem [{\citenamefont {F{\"o}rtsch}\ \emph {et~al.}(2013)\citenamefont
  {F{\"o}rtsch}, \citenamefont {F{\"u}rst}, \citenamefont {Wittmann},
  \citenamefont {Strekalov}, \citenamefont {Aiello}, \citenamefont {Chekhova},
  \citenamefont {Silberhorn}, \citenamefont {Leuchs},\ and\ \citenamefont
  {Marquardt}}]{foertsch2013}%
  \BibitemOpen
  \bibfield  {author} {\bibinfo {author} {\bibfnamefont {M.}~\bibnamefont
  {F{\"o}rtsch}}, \bibinfo {author} {\bibfnamefont {J.~U.}\ \bibnamefont
  {F{\"u}rst}}, \bibinfo {author} {\bibfnamefont {C.}~\bibnamefont {Wittmann}},
  \bibinfo {author} {\bibfnamefont {D.}~\bibnamefont {Strekalov}}, \bibinfo
  {author} {\bibfnamefont {A.}~\bibnamefont {Aiello}}, \bibinfo {author}
  {\bibfnamefont {M.~V.}\ \bibnamefont {Chekhova}}, \bibinfo {author}
  {\bibfnamefont {C.}~\bibnamefont {Silberhorn}}, \bibinfo {author}
  {\bibfnamefont {G.}~\bibnamefont {Leuchs}}, \ and\ \bibinfo {author}
  {\bibfnamefont {C.}~\bibnamefont {Marquardt}},\ }\href {\doibase
  10.1038/ncomms2838} {\bibfield  {journal} {\bibinfo  {journal} {Nature
  communications}\ }\textbf {\bibinfo {volume} {4}},\ \bibinfo {pages} {1818}
  (\bibinfo {year} {2013})}\BibitemShut {NoStop}%
\bibitem [{\citenamefont {Dao}\ \emph {et~al.}(2012)\citenamefont {Dao},
  \citenamefont {Aljunid}, \citenamefont {Maslennikov},\ and\ \citenamefont
  {Kurtsiefer}}]{dao2012}%
  \BibitemOpen
  \bibfield  {author} {\bibinfo {author} {\bibfnamefont {H.~L.}\ \bibnamefont
  {Dao}}, \bibinfo {author} {\bibfnamefont {S.~A.}\ \bibnamefont {Aljunid}},
  \bibinfo {author} {\bibfnamefont {G.}~\bibnamefont {Maslennikov}}, \ and\
  \bibinfo {author} {\bibfnamefont {C.}~\bibnamefont {Kurtsiefer}},\ }\href
  {\doibase http://dx.doi.org/10.1063/1.4739776} {\bibfield  {journal}
  {\bibinfo  {journal} {Review of Scientific Instruments}\ }\textbf {\bibinfo
  {volume} {83}},\ \bibinfo {pages} {083104} (\bibinfo {year}
  {2012})}\BibitemShut {NoStop}%
\bibitem [{\citenamefont {Kolchin}\ \emph {et~al.}(2008)\citenamefont
  {Kolchin}, \citenamefont {Belthangady}, \citenamefont {Du}, \citenamefont
  {Yin},\ and\ \citenamefont {Harris}}]{kolchin2008}%
  \BibitemOpen
  \bibfield  {author} {\bibinfo {author} {\bibfnamefont {P.}~\bibnamefont
  {Kolchin}}, \bibinfo {author} {\bibfnamefont {C.}~\bibnamefont
  {Belthangady}}, \bibinfo {author} {\bibfnamefont {S.}~\bibnamefont {Du}},
  \bibinfo {author} {\bibfnamefont {G.~Y.}\ \bibnamefont {Yin}}, \ and\
  \bibinfo {author} {\bibfnamefont {S.~E.}\ \bibnamefont {Harris}},\ }\href
  {\doibase 10.1103/PhysRevLett.101.103601} {\bibfield  {journal} {\bibinfo
  {journal} {Phys. Rev. Lett.}\ }\textbf {\bibinfo {volume} {101}},\ \bibinfo
  {pages} {103601} (\bibinfo {year} {2008})}\BibitemShut {NoStop}%
\bibitem [{\citenamefont {Srivathsan}\ \emph {et~al.}(2014)\citenamefont
  {Srivathsan}, \citenamefont {Gulati}, \citenamefont {Cer\`e}, \citenamefont
  {Chng},\ and\ \citenamefont {Kurtsiefer}}]{srivathsan2014}%
  \BibitemOpen
  \bibfield  {author} {\bibinfo {author} {\bibfnamefont {B.}~\bibnamefont
  {Srivathsan}}, \bibinfo {author} {\bibfnamefont {G.~K.}\ \bibnamefont
  {Gulati}}, \bibinfo {author} {\bibfnamefont {A.}~\bibnamefont {Cer\`e}},
  \bibinfo {author} {\bibfnamefont {B.}~\bibnamefont {Chng}}, \ and\ \bibinfo
  {author} {\bibfnamefont {C.}~\bibnamefont {Kurtsiefer}},\ }\href {\doibase
  10.1103/PhysRevLett.113.163601} {\bibfield  {journal} {\bibinfo  {journal}
  {Phys. Rev. Lett.}\ }\textbf {\bibinfo {volume} {113}},\ \bibinfo {pages}
  {163601} (\bibinfo {year} {2014})}\BibitemShut {NoStop}%
\bibitem [{\citenamefont {Heugel}\ \emph {et~al.}(2010)\citenamefont {Heugel},
  \citenamefont {Villar}, \citenamefont {Sondermann}, \citenamefont {Peschel},\
  and\ \citenamefont {Leuchs}}]{heugel2010}%
  \BibitemOpen
  \bibfield  {author} {\bibinfo {author} {\bibfnamefont {S.}~\bibnamefont
  {Heugel}}, \bibinfo {author} {\bibfnamefont {A.~S.}\ \bibnamefont {Villar}},
  \bibinfo {author} {\bibfnamefont {M.}~\bibnamefont {Sondermann}}, \bibinfo
  {author} {\bibfnamefont {U.}~\bibnamefont {Peschel}}, \ and\ \bibinfo
  {author} {\bibfnamefont {G.}~\bibnamefont {Leuchs}},\ }\href {\doibase
  10.1134/S1054660X09170095} {\bibfield  {journal} {\bibinfo  {journal} {Laser
  Physics}\ }\textbf {\bibinfo {volume} {20}},\ \bibinfo {pages} {100}
  (\bibinfo {year} {2010})},\ \Eprint {http://arxiv.org/abs/arXiv:1009.2365}
  {arXiv:1009.2365} \BibitemShut {NoStop}%
\bibitem [{\citenamefont {Bader}\ \emph {et~al.}(2013)\citenamefont {Bader},
  \citenamefont {Heugel}, \citenamefont {Chekhov}, \citenamefont {Sondermann},\
  and\ \citenamefont {Leuchs}}]{bader2013}%
  \BibitemOpen
  \bibfield  {author} {\bibinfo {author} {\bibfnamefont {M.}~\bibnamefont
  {Bader}}, \bibinfo {author} {\bibfnamefont {S.}~\bibnamefont {Heugel}},
  \bibinfo {author} {\bibfnamefont {A.~L.}\ \bibnamefont {Chekhov}}, \bibinfo
  {author} {\bibfnamefont {M.}~\bibnamefont {Sondermann}}, \ and\ \bibinfo
  {author} {\bibfnamefont {G.}~\bibnamefont {Leuchs}},\ }\href {\doibase
  10.1088/1367-2630/15/12/123008} {\bibfield  {journal} {\bibinfo  {journal}
  {New Journal of Physics}\ }\textbf {\bibinfo {volume} {15}},\ \bibinfo
  {pages} {123008} (\bibinfo {year} {2013})},\ \Eprint
  {http://arxiv.org/abs/arXiv:1309.6167 [physics.optics]} {arXiv:1309.6167
  [physics.optics]} \BibitemShut {NoStop}%
\bibitem [{\citenamefont {Palomaki}\ \emph {et~al.}(2013)\citenamefont
  {Palomaki}, \citenamefont {Harlow}, \citenamefont {Teufel}, \citenamefont
  {Simmonds},\ and\ \citenamefont {Lehnert}}]{palomaki2013}%
  \BibitemOpen
  \bibfield  {author} {\bibinfo {author} {\bibfnamefont {T.}~\bibnamefont
  {Palomaki}}, \bibinfo {author} {\bibfnamefont {J.}~\bibnamefont {Harlow}},
  \bibinfo {author} {\bibfnamefont {J.}~\bibnamefont {Teufel}}, \bibinfo
  {author} {\bibfnamefont {R.}~\bibnamefont {Simmonds}}, \ and\ \bibinfo
  {author} {\bibfnamefont {K.}~\bibnamefont {Lehnert}},\ }\href {\doibase
  10.1038/nature11915} {\bibfield  {journal} {\bibinfo  {journal} {Nature}\
  }\textbf {\bibinfo {volume} {495}},\ \bibinfo {pages} {210} (\bibinfo {year}
  {2013})}\BibitemShut {NoStop}%
\bibitem [{\citenamefont {Wenner}\ \emph {et~al.}(2014)\citenamefont {Wenner},
  \citenamefont {Yin}, \citenamefont {Chen}, \citenamefont {Barends},
  \citenamefont {Chiaro}, \citenamefont {Jeffrey}, \citenamefont {Kelly},
  \citenamefont {Megrant}, \citenamefont {Mutus}, \citenamefont {Neill},
  \citenamefont {O'Malley}, \citenamefont {Roushan}, \citenamefont {Sank},
  \citenamefont {Vainsencher}, \citenamefont {White}, \citenamefont {Korotkov},
  \citenamefont {Cleland},\ and\ \citenamefont {Martinis}}]{wenner2014}%
  \BibitemOpen
  \bibfield  {author} {\bibinfo {author} {\bibfnamefont {J.}~\bibnamefont
  {Wenner}}, \bibinfo {author} {\bibfnamefont {Y.}~\bibnamefont {Yin}},
  \bibinfo {author} {\bibfnamefont {Y.}~\bibnamefont {Chen}}, \bibinfo {author}
  {\bibfnamefont {R.}~\bibnamefont {Barends}}, \bibinfo {author} {\bibfnamefont
  {B.}~\bibnamefont {Chiaro}}, \bibinfo {author} {\bibfnamefont
  {E.}~\bibnamefont {Jeffrey}}, \bibinfo {author} {\bibfnamefont
  {J.}~\bibnamefont {Kelly}}, \bibinfo {author} {\bibfnamefont
  {A.}~\bibnamefont {Megrant}}, \bibinfo {author} {\bibfnamefont {J.~Y.}\
  \bibnamefont {Mutus}}, \bibinfo {author} {\bibfnamefont {C.}~\bibnamefont
  {Neill}}, \bibinfo {author} {\bibfnamefont {P.~J.~J.}\ \bibnamefont
  {O'Malley}}, \bibinfo {author} {\bibfnamefont {P.}~\bibnamefont {Roushan}},
  \bibinfo {author} {\bibfnamefont {D.}~\bibnamefont {Sank}}, \bibinfo {author}
  {\bibfnamefont {A.}~\bibnamefont {Vainsencher}}, \bibinfo {author}
  {\bibfnamefont {T.~C.}\ \bibnamefont {White}}, \bibinfo {author}
  {\bibfnamefont {A.~N.}\ \bibnamefont {Korotkov}}, \bibinfo {author}
  {\bibfnamefont {A.~N.}\ \bibnamefont {Cleland}}, \ and\ \bibinfo {author}
  {\bibfnamefont {J.~M.}\ \bibnamefont {Martinis}},\ }\href {\doibase
  10.1103/PhysRevLett.112.210501} {\bibfield  {journal} {\bibinfo  {journal}
  {Phys. Rev. Lett.}\ }\textbf {\bibinfo {volume} {112}},\ \bibinfo {pages}
  {210501} (\bibinfo {year} {2014})}\BibitemShut {NoStop}%
\bibitem [{\citenamefont {Liu}\ \emph {et~al.}(2014)\citenamefont {Liu},
  \citenamefont {Sun}, \citenamefont {Zhao}, \citenamefont {Zhang},
  \citenamefont {Loy},\ and\ \citenamefont {Du}}]{liu2014}%
  \BibitemOpen
  \bibfield  {author} {\bibinfo {author} {\bibfnamefont {C.}~\bibnamefont
  {Liu}}, \bibinfo {author} {\bibfnamefont {Y.}~\bibnamefont {Sun}}, \bibinfo
  {author} {\bibfnamefont {L.}~\bibnamefont {Zhao}}, \bibinfo {author}
  {\bibfnamefont {S.}~\bibnamefont {Zhang}}, \bibinfo {author} {\bibfnamefont
  {M.~M.~T.}\ \bibnamefont {Loy}}, \ and\ \bibinfo {author} {\bibfnamefont
  {S.}~\bibnamefont {Du}},\ }\href {\doibase 10.1103/PhysRevLett.113.133601}
  {\bibfield  {journal} {\bibinfo  {journal} {Phys. Rev. Lett.}\ }\textbf
  {\bibinfo {volume} {113}},\ \bibinfo {pages} {133601} (\bibinfo {year}
  {2014})}\BibitemShut {NoStop}%
\bibitem [{\citenamefont {Maiwald}\ \emph {et~al.}(2009)\citenamefont
  {Maiwald}, \citenamefont {Leibfried}, \citenamefont {Britton}, \citenamefont
  {Bergquist}, \citenamefont {Leuchs},\ and\ \citenamefont
  {Wineland}}]{maiwald2009}%
  \BibitemOpen
  \bibfield  {author} {\bibinfo {author} {\bibfnamefont {R.}~\bibnamefont
  {Maiwald}}, \bibinfo {author} {\bibfnamefont {D.}~\bibnamefont {Leibfried}},
  \bibinfo {author} {\bibfnamefont {J.}~\bibnamefont {Britton}}, \bibinfo
  {author} {\bibfnamefont {J.~C.}\ \bibnamefont {Bergquist}}, \bibinfo {author}
  {\bibfnamefont {G.}~\bibnamefont {Leuchs}}, \ and\ \bibinfo {author}
  {\bibfnamefont {D.~J.}\ \bibnamefont {Wineland}},\ }\href {\doibase
  10.1038/nphys1311} {\bibfield  {journal} {\bibinfo  {journal} {Nature
  Physics}\ }\textbf {\bibinfo {volume} {5}},\ \bibinfo {pages} {551} (\bibinfo
  {year} {2009})},\ \Eprint {http://arxiv.org/abs/arxiv:0810.2647}
  {arxiv:0810.2647} \BibitemShut {NoStop}%
\bibitem [{\citenamefont {Bell}\ \emph {et~al.}(1991)\citenamefont {Bell},
  \citenamefont {Gill}, \citenamefont {Klein}, \citenamefont {Levick},
  \citenamefont {Tamm},\ and\ \citenamefont {Schnier}}]{bell1991}%
  \BibitemOpen
  \bibfield  {author} {\bibinfo {author} {\bibfnamefont {A.~S.}\ \bibnamefont
  {Bell}}, \bibinfo {author} {\bibfnamefont {P.}~\bibnamefont {Gill}}, \bibinfo
  {author} {\bibfnamefont {H.~A.}\ \bibnamefont {Klein}}, \bibinfo {author}
  {\bibfnamefont {A.~P.}\ \bibnamefont {Levick}}, \bibinfo {author}
  {\bibfnamefont {C.}~\bibnamefont {Tamm}}, \ and\ \bibinfo {author}
  {\bibfnamefont {D.}~\bibnamefont {Schnier}},\ }\href@noop {} {\bibfield
  {journal} {\bibinfo  {journal} {Phys. Rev. A}\ }\textbf {\bibinfo {volume}
  {44}},\ \bibinfo {pages} {20} (\bibinfo {year} {1991})}\BibitemShut {NoStop}%
\end{thebibliography}


\end{document}